\documentclass[10pt,preprint]{aastex}
\received{}
\revised{}
\accepted{}
\shorttitle{Spectral Line Survey of Sgr B2(N-LMH)}
\shortauthors{Friedel et al.}

\begin{document}
\singlespace
\title{A Spectral Line Survey of Selected 3 mm bands\\
Toward Sagittarius~B2(N-LMH)\\
Using the NRAO 12 Meter Radio Telescope\\
and the BIMA Array \\
I. The Observational Data}

\author{D. N. Friedel\altaffilmark{1}, L. E. Snyder\altaffilmark{1}, B. E. Turner\altaffilmark{2} \and A. Remijan\altaffilmark{1}}
\altaffiltext{1}{Department of Astronomy, University of Illinois, Urbana, IL 61801\\email: friedel@astro.uiuc.edu, snyder@astro.uiuc.edu, aremijan@astro.uiuc.edu}
\altaffiltext{2}{National Radio Astronomy Observatory, Charlottesville, VA 22903\\email: bturner@nrao.edu}

\begin{abstract}
We have initiated a spectral line survey, at a wavelength of 3 millimeters, toward the hot molecular core
Sagittarius~B2(N-LMH). This is the first spectral line survey of the Sgr~B2(N) region utilizing data from both an
interferometer (BIMA Array) and a single-element radio telescope (NRAO 12 meter). In this survey, covering 3.6 GHz in
bandwidth, we detected 218 lines (97 identified molecular transitions, 1 recombination line, and 120 unidentified
transitions). This yields a spectral line density (lines per 100 MHz) of 6.06, which is much larger than any previous
3 mm line survey. We also present maps from the BIMA Array that indicate that most highly saturated species (3 or
more H atoms) are products of grain chemistry or warm gas phase chemistry. Due to the nature of this survey we are
able to probe each spectral line on multiple spatial scales, yielding information that could not be obtained by
either instrument alone.
\end{abstract}


\section{Introduction}
Sagittarius B2 is a massive star forming region at a distance of 7.1 kpc and within 300 pc of the
Galactic center \citep{reid88}. It is well known for its abundance of molecular species and has been the
target of many searches for complex molecules \citep[e.g.,][]{mehringer96,liu01,martin01} and several spectral
line surveys \citep[e.g.,][]{cummins86,turner89,turner91}. The region is made up of several individual sources.
Table~\ref{tab:sgrcoords} contains a list of the main sources in the region along with their coordinates. The
first column gives the source name. The second and third columns give the right ascension and declination of
each source, respectively. The dense molecular core toward Sgr~B2(N) has been called the ``Large Molecule Heimat''
(Sgr B2(N-LMH)) since it is the source of all of the large molecules observed in that region \citep{snyder94,mehringer96}.
Sgr~B2(N-LMH) has the most detected and identified molecular species and has the
strongest spectral emission lines of any source in the complex.\footnote{With the exception of
SO and SO$_{2}$ which are strongest toward Sgr~B2(M) \citep{sutton91}.} Within the 3~mm synthesized beam of the
Berkeley-Illinois-Maryland Association (BIMA) Array\footnote{Operated by the University of California,
Berkeley, the University of Illinois, and the University of Maryland with support from the National Science
Foundation.} in its C configuration toward Sgr~B2(N) there are at least four H{\sc{ii}} regions \citep{depree95}.
This is unlike Sgr~B2(M), which has over 21 H{\sc{ii}} regions in the same size beam \citep{gaume95,depree96}.

Spectral line surveys provide powerful tools to analyze the coupled dynamical and chemical evolution of molecular
clouds.  In a survey, many lines of a single species are observed at similar sensitivities, diminishing the errors
associated with derived quantities. For example, a particular problem arises in deriving accurate abundances due to
overlapping identifications of molecular species. Surveys provide the best means of avoiding this problem by
revealing intensity changes in lines of a given species.  The first line surveys of the Sgr~B2 source complex were
aimed at Sgr~B2(OH). Sgr~B2(OH) is an OH and H$_{2}$O maser source (${\sim}30\arcsec$ south of Sgr~B2(M)),
which is not a main site of molecular concentration in the region \citep{snyder91} and furthermore has
little or no millimeter continuum emission (${\le}0.1$ Jy beam$^{-1}$) \citep{kuan94,kuan96}.
However, the main beams of these surveys were large enough to encompass Sgr~B2(M) and in the case of
\citet{cummins86}, possibly even Sgr~B2(N) at some frequencies. Thus, many of the molecular transitions detected
by these surveys may not have come from Sgr~B2(OH) but from Sgr~B2(M) or Sgr~B2(N). More recent surveys covering
330-355 GHz \citep{sutton91} and 218-263 GHz \citep{nummelin98, nummelin00} did use Sgr~B2(N) and Sgr~B2(M) as
the primary sources. Figure~\ref{fig:beams} shows the pointing centers and average beams from this and previous
spectral line surveys of the Sgr~B2 region. The five cores of Sgr~B2 are labeled (North, Northwest, Main, South,
and OH). The beams from this survey are in bold face centered on Sgr~B2(N). The dashed beams around Sgr~B2(N),
Sgr~B2(M), and Sgr~B2(NW) are from \citet{nummelin98,nummelin00}. The grey beams around Sgr~B2(N) and Sgr~B2(M)
(2 beams) are from \citet{sutton91}. The beams centered on Sgr~B2(OH) are from \citet{turner89,turner91} and
\citet{cummins86}.

This survey uses both the
BIMA Array (ten 6.1 meter elements) and the single-element NRAO 12~meter\footnote{The National Radio Astronomy
Observatory is a facility of the National Science Foundation operated under cooperative agreement by Associated
Universities, Inc.} radio telescope. Unlike previous surveys that only used single-element telescopes, we are
able to sample several spatial scales of this source.  A single-element telescope is best for detecting large
scale, extended emission, but poor at detecting compact sources due to beam dilution.  An array is insensitive
to structures larger than an angular size determined by the minimum spacing of its elements, and thus resolves
out large scale structure.  For the BIMA Array, this is $\sim$10 times the size of the synthesized beam
\citep{wright96}\footnote{BIMA Memoranda Series \citep{wright96} is available at: http://bima.astro.umd.edu/memo/memo45.ps.}.
This feature is very important in astrochemistry because the location of a particular molecular species may
indicate its formation mechanism. Molecules formed by low temperature gas phase reactions will be primarily
extended in emission. On the other hand, molecules formed on grains, while they may form anywhere in the cloud,
will evaporate from the grains and be detected in the warmer compact cores of the sources \citep{cas93}.
Similarly, molecules formed by warm temperature gas phase reactions will also be detected in the warmer cores.
Thus, a single element telescope would be most sensitive to molecules formed in extended low temperature gas
clouds, while an array would be most sensitive to molecules formed in warm cores (either in warm gas phase
reactions or initially on a grain and then desorbed from the surface).

Our spectral line survey is the first to use both interferometric data and single-element telescope data from
Sgr~B2(N-LMH) and is also the first 3 mm survey of this source. Here we present the results of a line
survey towards Sgr B2(N-LMH) of six 600 MHz wide spectral bands (Figure~\ref{fig:cover}). All bands were observed
with both the BIMA Array and the NRAO 12 meter radio telescope.

\section{Observations and Data}
The pointing center of our survey was Sgr~B2(N-LMH)\footnote{While this is the pointing center of all observations,
the beams of both telescopes also encompassed other sources within Sgr~B2(N).}. The spectral bands surveyed
(Figure~\ref{fig:cover}) were centered on 86.2, 86.8, 90.025, 106.58, 108.5 and 110.2 GHz.\footnote{While the BIMA
Array observed a wider spectral band than the 12 meter telescope in several instances, these data are not being
presented in this paper since there is not 12 meter data to compare it to, but will be presented in future papers.}
We chose these bands because each contains at least one previously identified molecular transition. In most cases,
the spectral bands contained many well known molecular species including silicon monoxide (SiO), sulfur monoxide
(SO), deuterated ammonia (NH$_2$D), methyl formate (HCOOCH$_3$, MeF), cyanogen ($^{13}$CN), carbon monoxide
($^{13}$CO), methyl cyanide (CH$_3$CN) and formamide (NH$_2$CHO).

\subsection{NRAO 12 Meter Telescope Observations}
The NRAO 12 meter telescope observations were carried out in 2000 July. The spectra observed with the 12~meter
telescope utilized the dual-channel SIS mixer receiver operating in a single-sideband mode with the image
sideband rejected at a typical level of 20 dB. The MAC spectrometer was used in its 600 MHz bandwidth mode with
a channel spacing of 195 kHz per channel. This gives a frequency resolution of 391 kHz per channel, owing to
internal Hanning weighting, and a velocity resolution of $\sim$1.4, 1.4, 1.3, 1.1, 1.1 and 1.1 km s$^{-1}$ for the
86.2, 86.8, 90.025, 106.58, 108.5 and 110.2 GHz bands.  The two receivers provided orthogonal linear
polarizations and integrated on source for 1.75-2.5 hours. Calibration of the intensity scale made use of the 
chopper wheel method, with the resulting data on the $T^*_A$ scale. This was converted to the $T^*_R$ scale by 
applying the main beam efficiency, which varies between 0.68 at 86 GHz and 0.62 at 110 GHz, and is accurate to 
within $\sim$5\% \citep{kutner81}. This scale corrects for atmospheric
extinction and telescope spillover losses, but not for error-beam losses or the forward beam coupling to the
source. Data were taken using position switching with the reference position 30$\arcmin$ west in azimuth in
order to avoid any extended structure. The passbands were calibrated by baseline fitting to the data with a 
polynomial of order 4 or less. The passbands were checked before and after calibration for spurious features;
 none were found. The 12 meter telescope data were analyzed with the UniPOPS software package
\citep{salt95}\footnote{UniPOPS information is available at: http://info.gb.nrao.edu/$\sim$rmaddale/140ft/unipops/unipops\_toc.html.}.

\subsection{BIMA Array Observations}
Observations with the BIMA Array were carried out in C configuration (minimum baseline of 8 meters and maximum
baseline of 61 meters) between 2001 March and 2002 April. The array was operating in cross-correlation mode
(double sideband) with a sideband rejection of better than 20 dB. The correlator configuration was
four 50 MHz spectral windows set side by side with the edges overlapping by 3 channels, giving an effective
bandwidth of just under 200 MHz in each sideband. Each window had 128 channels resulting in a spectral resolution
of 390 kHz per channel and a velocity resolution of $\sim$1.4, 1.4, 1.3, 1.1, 1.1 and 1.1 km s$^{-1}$ for the
86.2, 86.8, 90.025, 106.58, 108.5 and 110.2 GHz bands. This set of 4 windows was set up to start at the lower
end of each of the 600 MHz wide spectral bands and integrated on source for 50 minutes. Next, the observing
frequency was shifted by 197 MHz and the integration was continued. This process was repeated until the entire
600 MHz band was observed. Neptune and Mars were used as flux density calibrators and 1733-130 was used to
calibrate the antenna based gains. The absolute amplitude calibration of 1733-130 from the flux density calibrators
is accurate to within $\sim$20\%. The passbands were automatically calibrated online during data acquisition.\footnote{A 
technical description of this can be found at: http://astron.berkeley.edu/$\sim$plambeck/technical.html.} 
In the past this method has been quite satisfactory and has not generated spurious features.
The BIMA Array data were calibrated and imaged using the MIRIAD software package \citep{sault95}.

\subsection{Data}
Table~\ref{tab:rms} contains a list of beam sizes and rms
noise levels for the observations. The first column lists the central frequency of the observed band (GHz). The
second and third columns list the synthesized beam size (arcsec) and the 1$\sigma$ channel rms noise level (Jy
beam$^{-1}$) for the BIMA Array observations, which have been averaged over all windows. The last two columns list
the FWHM beam size (arcsec) and the 1$\sigma$ channel rms noise level (Jy beam$^{-1}$) of the 12 meter telescope observations.
The 1$\sigma$ rms noise levels from the BIMA Array were determined by taking the rms over all channels from a spatial region
that was free from continuum emission. The 1$\sigma$ rms noise levels from the 12 meter telescope were determined by
calculating the rms in sections of line free channels.

The spectra are presented in Figures~\ref{fig:spec}(a-l). In order to derive the frequency scale for the BIMA Array, 
$V_{LSR}$=64 km s$^{-1}$ was used \citep{mehringer96}. For the 12 meter telescope $V_{LSR}$=65 km s$^{-1}$ was used 
because it is the value in a standard 12 meter catalog. In the figures, the two upper panels show BIMA Array data, 
and the bottom panel shows 12 meter telescope data. The top panel in each figure shows BIMA Array data which are Hanning 
weighted to bring out weaker features by reducing the noise level. The middle panel in each figure shows the BIMA Array data which are 
unsmoothed (not Hanning weighted) in order to preserve narrow features which are also seen in the 12 meter telescope data. 
The 12 meter telescope data were automatically Hanning smoothed as part of the internal data acquisition routine. The spectral 
resolution of the unsmoothed BIMA Array data match that of the smoothed 12 meter telescope data to within 1 kHz.
The 12 meter telescope intensity scale was converted from $T^*_R$ to Jy~beam$^{-1}$ using
30.5, 30.6, 30.8, 31.6, 31.8, 32.0 Jy~K$^{-1}$ as conversion factors for the 86.2, 86.8, 90.025, 106.58, 108.5
and 110.2 GHz bands, respectively. The dashed lines on the BIMA Array spectra denote spectral window edges
and the ``I'' bars, in both the 12 meter telescope and BIMA Array data (unsmoothed), denote 1$\sigma$ rms
noise levels. The threshold for defining a line is as follows: it must have a signal-to-noise of at least 
3$\sigma$ and have a line width of at least 4 km s$^{-1}$ (unless a notable feature is seen at the same frequency 
with both instruments). Unidentified lines labeled with ``Q'' are questionable detections due to their narrow line width.

Table~\ref{tab:summ} summarizes the molecular detections of the survey. The first column lists
the molecule; the second and third columns list the number of detected transitions by the BIMA Array and the
12 meter telescope, respectively.  The fourth column lists the reference for the molecular information for
each species. Table~\ref{tab:mol} lists the molecular transitions grouped by molecule. The first column lists
the rest frequency of the transition, along with the 2$\sigma$ standard deviation. The second column lists the
quantum numbers of the transition. The third and fourth columns list the intensity and line width\footnote{Intensities 
and line widths were obtained by a Gaussian least squares fit to the line profiles.},
with the 2$\sigma$ standard deviation, of each line observed by the BIMA Array. Fits to the BIMA Array data
were on the unsmoothed spectra. If the transition was not detected, a 1$\sigma$ upper limit has been set for the intensity.
The fifth and sixth columns list the intensity and line width, with the 2$\sigma$ standard deviation, of each
line observed by the 12 meter telescope. In a single case from the BIMA Array, the blend of MeF and
NH$_2$D at 85.92 GHz (see Figure~\ref{fig:spec}a), the MeF lines were modeled using the fits to the other detected 
MeF transitions, in order to obtain a satisfactory fit to all components. The seventh column lists the upper
energy level of each transition. The eighth column lists the product of the line strength and the square of the
relevant dipole moment. Table~\ref{tab:U} lists the detected unidentified (U) lines. The first column lists the
rest frequency (based on the LSR velocity of Sgr~B2(N-LMH)).  The second and third columns list the peak
intensity and line width, with the 2$\sigma$ standard deviation, of each line from the BIMA Array.  The fourth
and fifth columns list the peak intensity and line width, with the 2$\sigma$ standard deviation, of each line
from the 12 meter telescope.  If a line was undetected by one of the telescopes, a 1$\sigma$ upper limit has
been set for the intensity. In order to find potential identifications for the U~lines, we searched the JPL
database \citep{jpl}, The Cologne Database for Molecular Spectroscopy \citep{cologne}, and the Lovas List, a
compilation of observed molecular transitions from the ISM \citep{lovas03}. Potential identifications were also
given by F.\ J.\ Lovas (private communication, hereafter FJL) and
J.~C.~Pearson (private communication, hereafter JCP). The sixth column gives these potential identifications,
which were made based on the rest frequency of the transition. Any transition having a 2$\sigma$ frequency
standard deviation greater than 1 MHz (3 km s$^{-1}$ at 3 mm) or any molecule containing astrophysically unlikely 
elements was excluded. Three of these transitions are
labeled as potentially coming from NH$_2$CH$_2$COOH-I (glycine), although the likelihood of this identification
is small (an explanation is given with each label). There are 21 U~lines listed in 
Table~\ref{tab:U} (denoted with footnote {\it a}) which do not meet both of the threshold criteria, 15 from the 
BIMA Array and 6 from the 12 meter telescope. While the origin of these lines is unknown, those detected by the BIMA 
Array are most likely coming from the warmest and most compact parts of the core and could be vibrationally excited 
transitions of unidentified species. Those detected by the 12 meter telescope are most likely coming from the extended 
cool gas around the core. These lines are not included in any of the line counts or statistics.

\section{Results}
\subsection{Statistics}
In this survey, we detected a total of 218 lines (97 identified molecular transitions from 18 molecular
species and 16 isotopomers, 1 recombination line, and 120 U~lines). The BIMA Array detected 199 lines (91 from
15 molecular species and 16 isotopomers, and 108 U~lines) and the 12 meter telescope detected 116 lines (74
from 17 molecular species and 15 isotopomers, 1 recombination line, and 41 U~lines). Note that the number of
detected lines does not include hyperfine transitions. Of the lines detected by the BIMA Array, 21 are
previously undetected transitions of known molecules, 102 are previously undetected U~lines, and 101 (23 transitions of known molecules and 79
U~lines) were only detected by the BIMA Array. The 12 meter telescope detected 12 previously undetected
transitions of known molecules, 37 new U~lines, and 19 (7 transitions of known molecules and 12 U~lines) were only detected by the 12 meter telescope.

The above line counts yield a spectral line density (average number of lines per 100 MHz) of 6.06 from all
observed lines (2.72 from identified lines and 3.34 from U~lines). This is a large increase over previous
3~mm surveys of the Sgr~B2 region. Table~\ref{tab:stats} compares the line counts and densities from this
survey to previous 3~mm surveys of this source and from \citet{lovas03}. The first column gives the source of
the data. The second and third columns give the line counts and densities (from the same frequency bands
covered in this survey) of identified transitions from each source. The fourth and fifth columns give the line
counts and density of U~lines from each source. The sixth and seventh columns give the total line counts and
densities from each source. The total line density from this survey is greater than from either of the
previous line surveys and from \citet{lovas03}. The largest increase is the number of U~lines detected; a
total of $\sim$55\% of the lines we detected are unidentified. This demonstrates that with higher sensitivity
and smaller beams,  more spectral features will be detected.

Due to the relatively large line widths (typically 5-10 km s$^{-1}$) and the large number of molecular species in the
region, many spectral features are unresolved.  In cases of very strong emission from a single transition, nearby
weaker transitions may not have been observed. Thus if a specific transition was not detected, it may either currently
lie below our detection limit or may be blended in the wings of stronger lines.

\subsection{Maps}
Figure~\ref{fig:mapa} shows maps from the BIMA Array toward Sgr~B2(N-LMH) of NH$_2$CHO (formamide), C$_2$H$_5$OH (ethanol, EtOH),
SO, and H$^{13}$CN.
In each map, the continuum is mapped in grey scale with the transition averaged over the FWHM of the line mapped in
contours. The central continuum peak is Sgr~B2(N) and the lower continuum peak is Sgr~B2(M). The BIMA Array and 12 meter telescope
spectra of each mapped transition are below their respective map. The vertical axis is intensity in Jy beam$^{-1}$
and the horizontal axis is frequency in GHz. The synthesized beam is given in the lower left corner of each map. The
numbers on the grey scale wedge are in units of Jy beam$^{-1}$.

\section{Discussion}
The higher ($>$90 K) upper state energy (E$_u$) transitions of all species will map toward the
core since these transitions are only populated to detectable levels in the warmer temperature regions. But the
lower E$_u$ transitions can be excited in both the warm core and the cooler surrounding gas. Thus, the lower E$_u$ transitions
trace a truer distribution of each molecular species and are an indicator of that molecule's formation mechanism.
Since hydrogenation of atoms heavier than H and molecules happens with very high efficiency on grains
\citep{watson72}, all transitions of highly saturated (3 or more H atoms) molecules should map toward the core where
the molecules are desorbing from the grain surfaces. Similarly, any molecule formed by reactions by ions or other heavier atoms
and molecules on grain surfaces should map toward the core. We also expect any molecules formed in warm gas phase reactions
will map primarily toward the hot core since this is where the temperatures are warm enough for the reactions to proceed.
The lower E$_u$ transitions of species which are produced by cool gas phase reactions should have a large extended component
in their distribution, since these reactions do not require the higher temperatures of the core to proceed.

Figure \ref{fig:mapa}(a) shows the BIMA Array map of the 106.541 GHz, 5$_{2,3}$-4$_{2,2}$ transition of NH$_2$CHO which has
an E$_u$=27.2 K. The contours are 3, 6, 9, 12, 15, 18, and 21$\sigma$.  The average (across the FWHM of the line) flux
detected by the BIMA Array is 6.27(29)\footnote{Errors quoted are 1$\sigma$ rms noise level for the associated transition.}
Jy beam$^{-1}$ while the average flux detected by the 12 meter telescope is 8.61(63) Jy beam$^{-1}$. This map and the
difference in flux indicate that NH$_2$CHO has a compact distribution toward the core with few extended components and most
likely is the product of grain chemistry or warm gas phase chemistry. The observations by \citet{shilke91} agree with this
view of the NH$_2$CHO distribution. With a 17$\arcsec$ beam they found unresolved NH$_2$CHO emission toward Sgr~B2(N) with
much weaker emission toward the Sgr~B2(M) region. This map is indicative of most of the highly saturated species we have observed.

One of the exceptions is shown in Figure \ref{fig:mapa}(b), which is the BIMA Array map of the 90.117 GHz, 4$_{1,4}$-3$_{0,3}$
transition of EtOH with an E$_u$=9.4 K. The contours are -2, -1, 1, 3, and 4.5$\sigma$. This transition has peaks and
valleys throughout the map and has an average flux of 0.60(15) Jy beam$^{-1}$ from the BIMA Array and 8.20(68) Jy beam$^{-1}$
from the 12 meter telescope. The 12 meter telescope spectrum has a much wider line as well as several more velocity components
than the BIMA Array spectrum. This indicates that EtOH has a highly extended distribution and suggests a cooler gas phase
formation mechanism. The 106.775 GHz, 9$_{1,8}$-8$_{2,7}$~AA (E$_u$=43.4 K) transition of (CH$_3$)$_2$O (dimethyl ether)
(see Figure~\ref{fig:spec}h),
the other exception, maps similarly to this transition of EtOH. While there is no great intensity difference between the BIMA Array
spectrum and 12 meter telescope spectrum of this transition, the 12 meter telescope spectrum has wider lines and several
velocity components, indicating an extended distribution and a cool gas phase formation mechanism.

Figure \ref{fig:mapa}(c) shows the BIMA Array map of the 86.093 GHz, 2$_2$-1$_1$ transition of SO which has an E$_u$=19.3 K.
The contours are -6, -3, 3, 6, 9, 12, 15, 18, 21, and 24$\sigma$. The average flux detected by the BIMA Array is 6.69(28)
Jy beam$^{-1}$ and the average flux detected by the 12 meter telescope is 16.30(61) Jy beam$^{-1}$. This transition show
the distinction between sulfurated species and highly saturated organic species in the region. It has strong peaks toward
both Sgr~B2(N-LMH) and Sgr~B2(M) as well as strong peaks and valleys throughout the rest of the map. While most of the
flux is coming from the hot core regions the flux difference between the two instruments along with the map indicated that
SO has significant extended structure.

Figure \ref{fig:mapa}(d) shows the BIMA Array map of the 86.340 GHz, 1$_{0,1}$-0$_{0,0}$ transition of H$^{13}$CN which has an
E$_u$=4.0 K. The contours are -14, -12, -9, -6, -3, and 3$\sigma$. The map of this transition is indicative of the other
species which also show strong absorption. This map, across the FWHM of the absorption, does not show any of the emission
components seen in the 12 meter telescope spectrum. This is because these components are very extended and are being resolved
out by the BIMA Array.

These maps show that the highly saturated species\footnote{No conclusion about the spatial distribution of CH$_3$OH can be made
from our data since both detected transitions have a higher E$_u$ of 102 K.} such as EtCN, CH$_3$CN, MeF, NH$_2$CHO, and VyCN map
toward the Sgr~B2(N-LMH) core, indicating that these species are products of grain or warm gas phase chemistry. Contrasting
this, two other highly saturated species, EtOH and (CH$_3$)$_2$O, show highly extended distributions, indicating a cooler gas phase formation mechanism.

\section{Summary}
We have presented the spectra and line fits from the first 3~mm spectral line survey using both an interferometer
and a single element radio telescope of the Sgr~B2(N-LMH) hot molecular core. We detected 218 lines (97 identified
molecular transitions from 18 molecular species and 16 isotopomers, 1 recombination line, and 120 unidentified
transitions). The survey has a spectral line density of 6.06 lines per 100 MHz from all observed lines (2.72 from
identified lines and 3.34 from U~lines) which is much greater than any previous 3~mm survey of this source. The
high density of U~lines we detected ($\sim$55\% of all detected lines) indicates that as Sgr~B2(N-LMH) is observed
with increasing sensitivity and smaller beams, more spectral features will be seen.

Due to the nature of this survey we are able to probe each spectral line on multiple spatial scales, yielding
information that could not be obtained by either instrument alone. From this spatial information we can constrain
potential formation mechanisms for the identified molecular species. Products of cool gas phase reactions will
show extended distributions while products of warm gas phase reactions or grain surface reactions will show compact
distributions. The BIMA Array maps show that the observed highly saturated species such as EtCN, CH$_3$CN, MeF,
NH$_2$CHO, and VyCN show compact emission from the Sgr~B2(N-LMH) core, indicating a grain chemistry
or warm gas phase chemistry formation mechanism. Contrasting this, two other highly saturated species, EtOH and (CH$_3$)$_2$O,
show very extended distributions, indicating a cooler gas phase chemistry formation mechanism. A detailed analysis of
selected data sets, including temperature determinations, column densities, and comparison of results with existing
chemical models will be presented in a subsequent paper.

\acknowledgments
We thank D.~S.~Meier for many helpful discussions. F.~J.~Lovas, J.~C.~Pearson, and J.~M.~Hollis helped us identify 
U~lines. We thank A. Apponi for help with information on the 12 meter telescope and J.~R.~Forster for help with information 
on the BIMA Array. We also thank an anonymous referee for many helpful comments which improved this manuscript. 
We acknowledge support from the Laboratory for Astronomical Imaging at the University of Illinois, NSF AST 99-81363, 
and NSF AST 02-28953.

\clearpage
\figcaption{Pointing centers and average beams from this and previous spectral line surveys of the Sgr~B2 region. The five cores of Sgr~B2 are labeled (North, Main, South, Northwest, and OH). The beams from this survey are in bold face centered on
Sgr~B2(N). The dashed beams around Sgr~B2(N), Sgr~B2(M), and Sgr~B2(NW) are from \protect{\citet{nummelin98,nummelin00}}. The
grey beams around Sgr~B2(N) and Sgr~B2(M) (2 beams) are from \protect{\citet{sutton91}}. The beams centered on Sgr~B2(OH) are
from \protect{\citet{turner89,turner91}} and \protect{\citet{cummins86}}. \label{fig:beams}}

\figcaption{Frequency coverage of the line survey. 12 meter telescope coverage is in black. BIMA Array coverage is in grey. Some
frequencies are covered by the BIMA Array and not the 12 meter telescope due to the dual sidebands of the BIMA Array receiver and
correlator systems. Only frequencies which have been observed by both instruments are presented in this paper.\label{fig:cover}}

\figcaption{The top spectra are from the BIMA Array. The vertical dashed lines denote where two spectral windows meet.
The bars next to the window edges denote the rms noise level of each window. The bottom spectra are from the NRAO 12
meter and match the BIMA Array frequency range. The 12 meter telescope intensity scale was converted from $T^*_R$ to
Jy~beam$^{-1}$ using 30.5,
30.6, 30.8, 31.6, 31.8, 32.0 Jy~K$^{-1}$ as conversion factors for the 86.2, 86.8, 90.025, 106.58, 108.5 and 110.2 GHz
bands, respectively. Detected transitions of known lines are noted, as are lines from unknown transitions (denoted as
U).\label{fig:spec}}

\figcaption{Maps from the BIMA Array toward Sgr~B2(N-LMH) of 4 representative molecular species.
In each map the continuum is mapped in grey scale with the transition, averaged over the FWHM of the line, mapped in
contours. The central continuum peak is Sgr~B2(N-LMH) and the lower continuum peak is Sgr~B2(M). The BIMA Array and
12 meter telescope spectra of each
mapped transition is below their respective maps. The vertical axis is intensity in Jy beam$^{-1}$
and the horizontal axis is frequency in GHz. The synthesized beam is given in the lower left corner of each map.
The numbers on the grey scale wedge are in Jy beam$^{-1}$. (a) Map of the 106.541 GHz, 5$_{2,3}$-4$_{2,2}$ transition of
NH$_2$CHO (formamide). The contours are 3, 6, 9, 12, 15, 18, and 21$\sigma$. (b) The 90.117 GHz, 4$_{1,4}$-3$_{0,3}$
transition of C$_2$H$_5$OH (ethanol, EtOH). The contours are -2, -1, 1, 3, and 4.5$\sigma$. (c) Map of the 86.093 GHz,
2$_2$-1$_1$ transition of SO. The contours are -6, -3, 3, 6, 9, 12, 15, 18, 21, and 24$\sigma$. (d) Map of the 86.340 GHz,
1$_{0,1}$-0$_{0,0}$ transition of H$^{13}$CN. The contours are -14, -12, -9, -6, -3, and 3$\sigma$.\label{fig:mapa}}

\clearpage
\begin{figure}
\epsscale{0.9}
\plotone{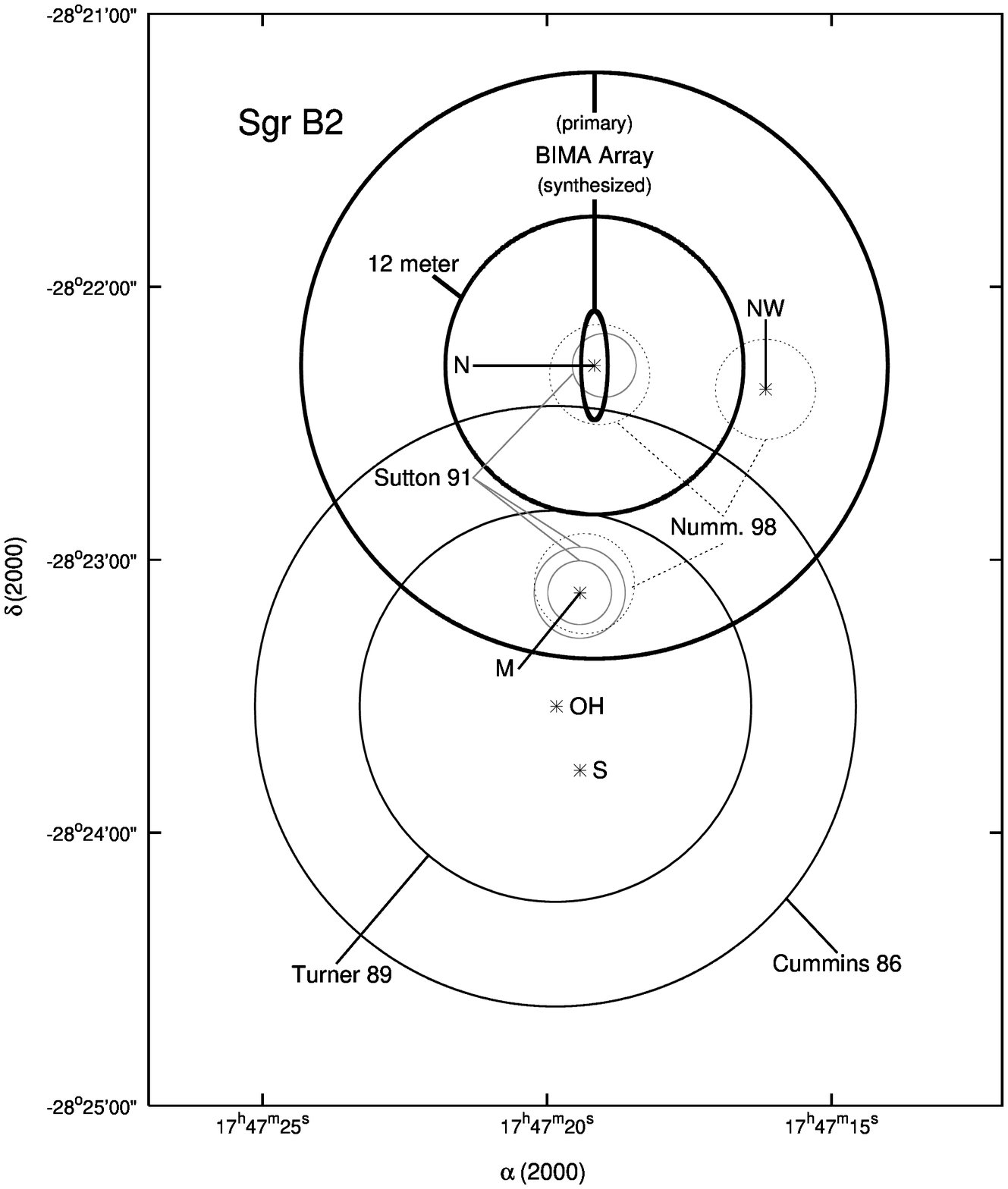}
\centerline{Figure \ref{fig:beams}}
\end{figure}
\clearpage
\begin{figure}
\epsscale{0.8}
\plotone{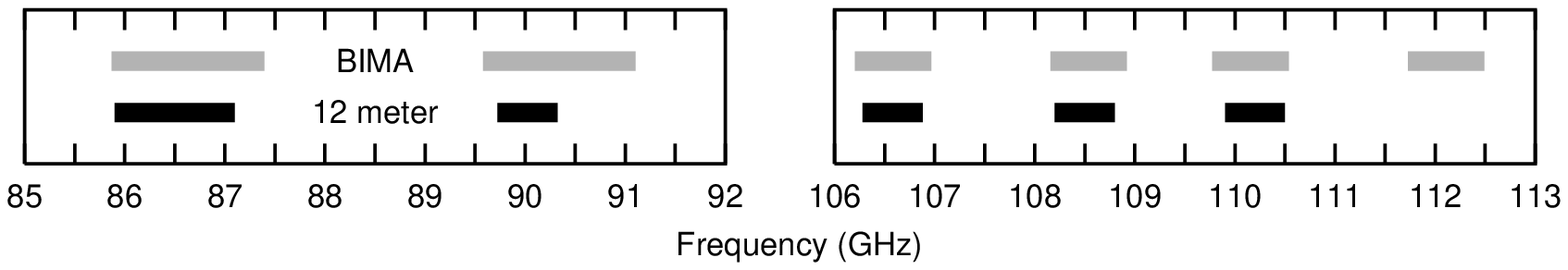}
\centerline{Figure \ref{fig:cover}}
\end{figure}
\clearpage
\begin{figure}
\epsscale{1.1}
\plotone{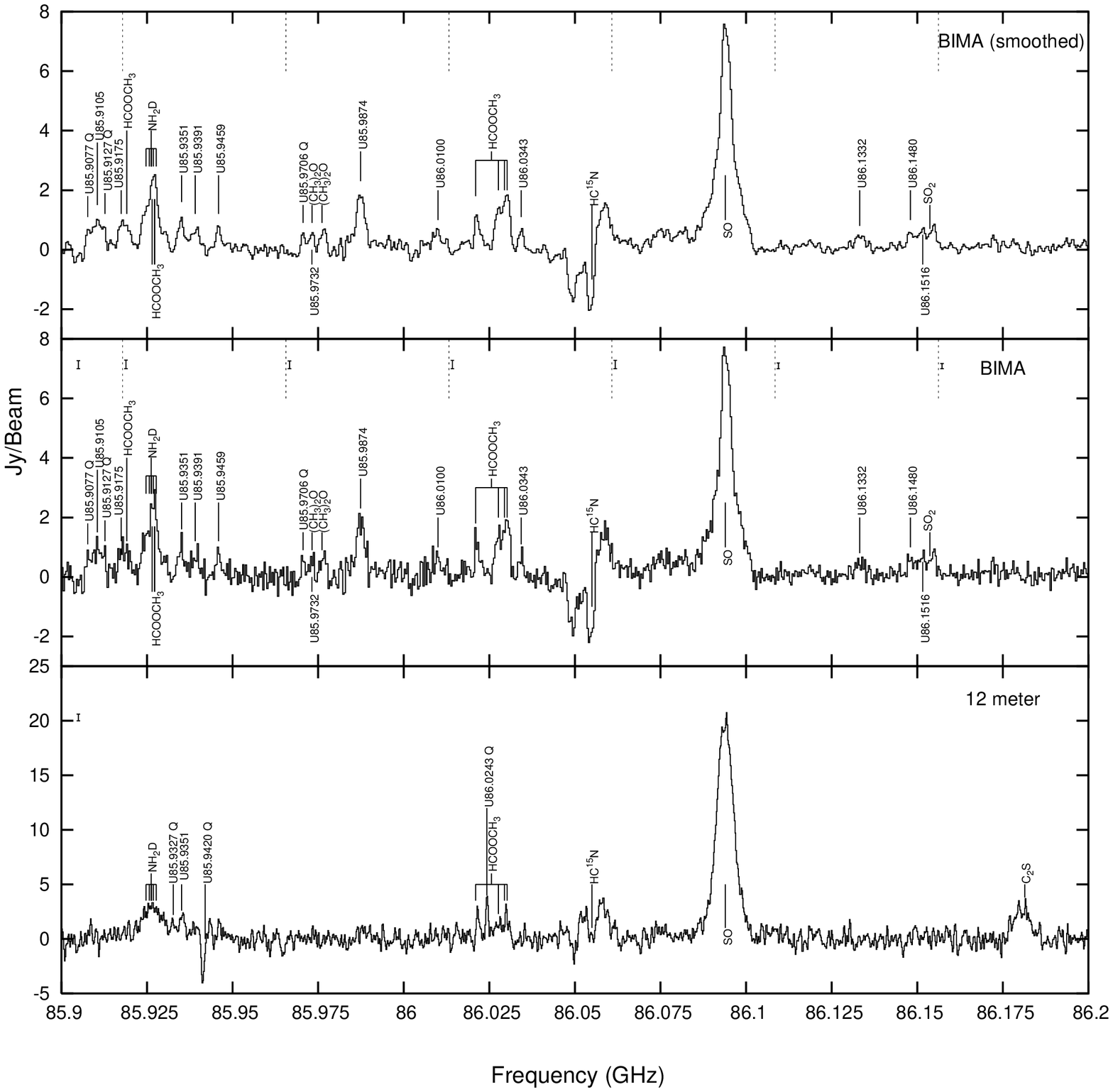}
\centerline{Figure \ref{fig:spec}a}
\end{figure}
\clearpage
\begin{figure}
\epsscale{1.1}
\plotone{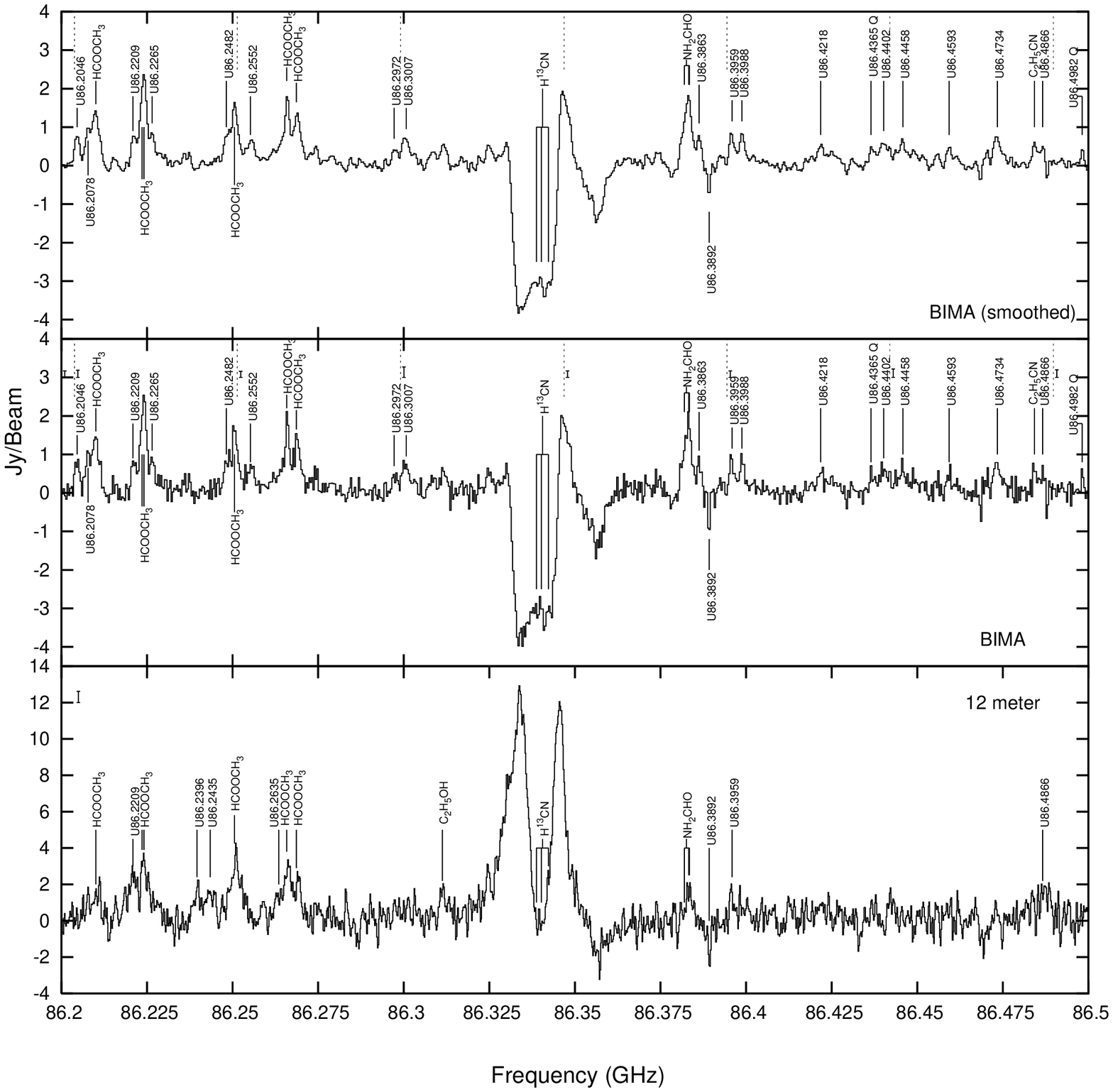}
\centerline{Figure \ref{fig:spec}b}
\end{figure}
\clearpage
\begin{figure}
\epsscale{1.1}
\plotone{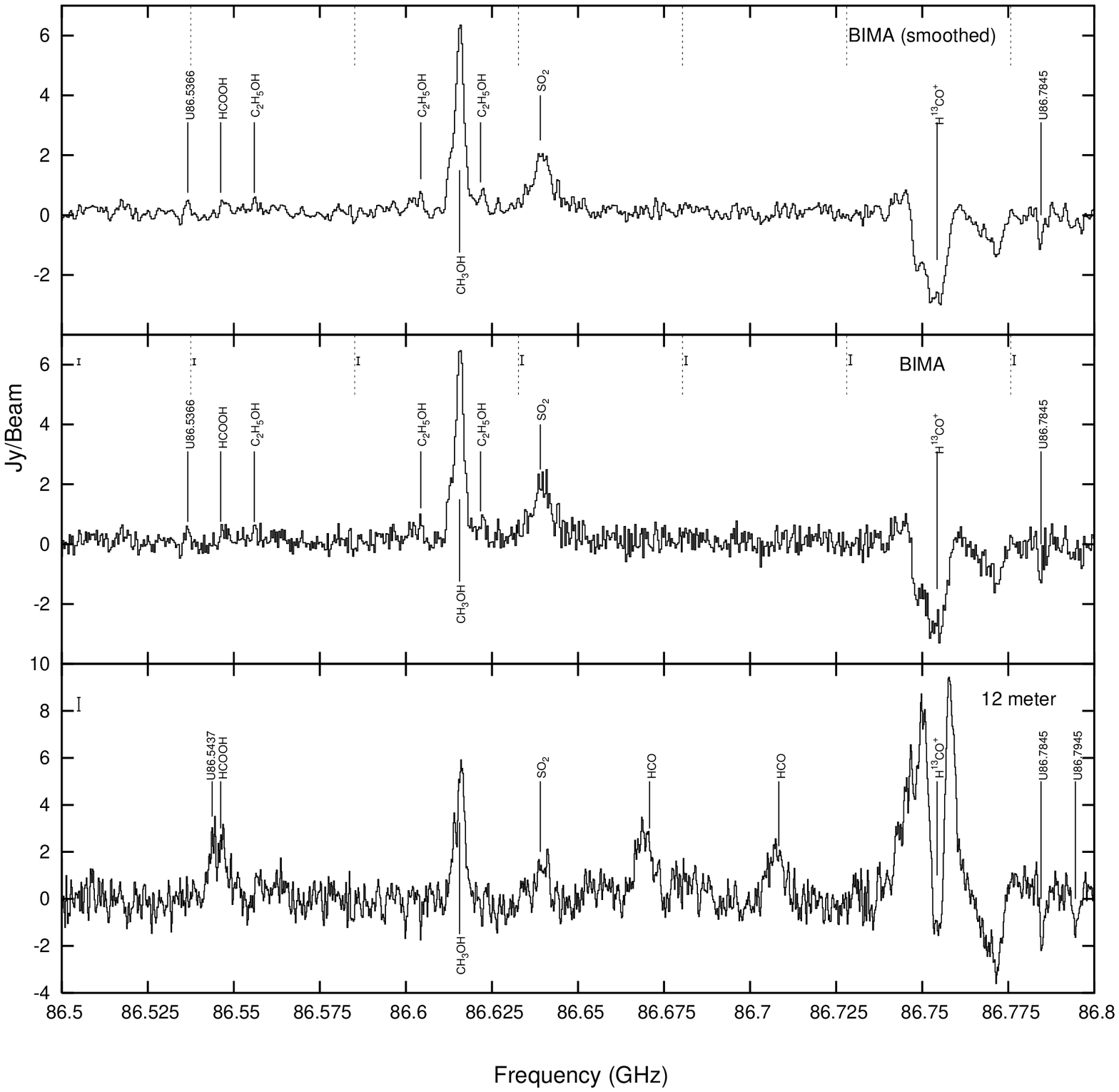}
\centerline{Figure \ref{fig:spec}c}
\end{figure}
\clearpage
\begin{figure}
\epsscale{1.1}
\plotone{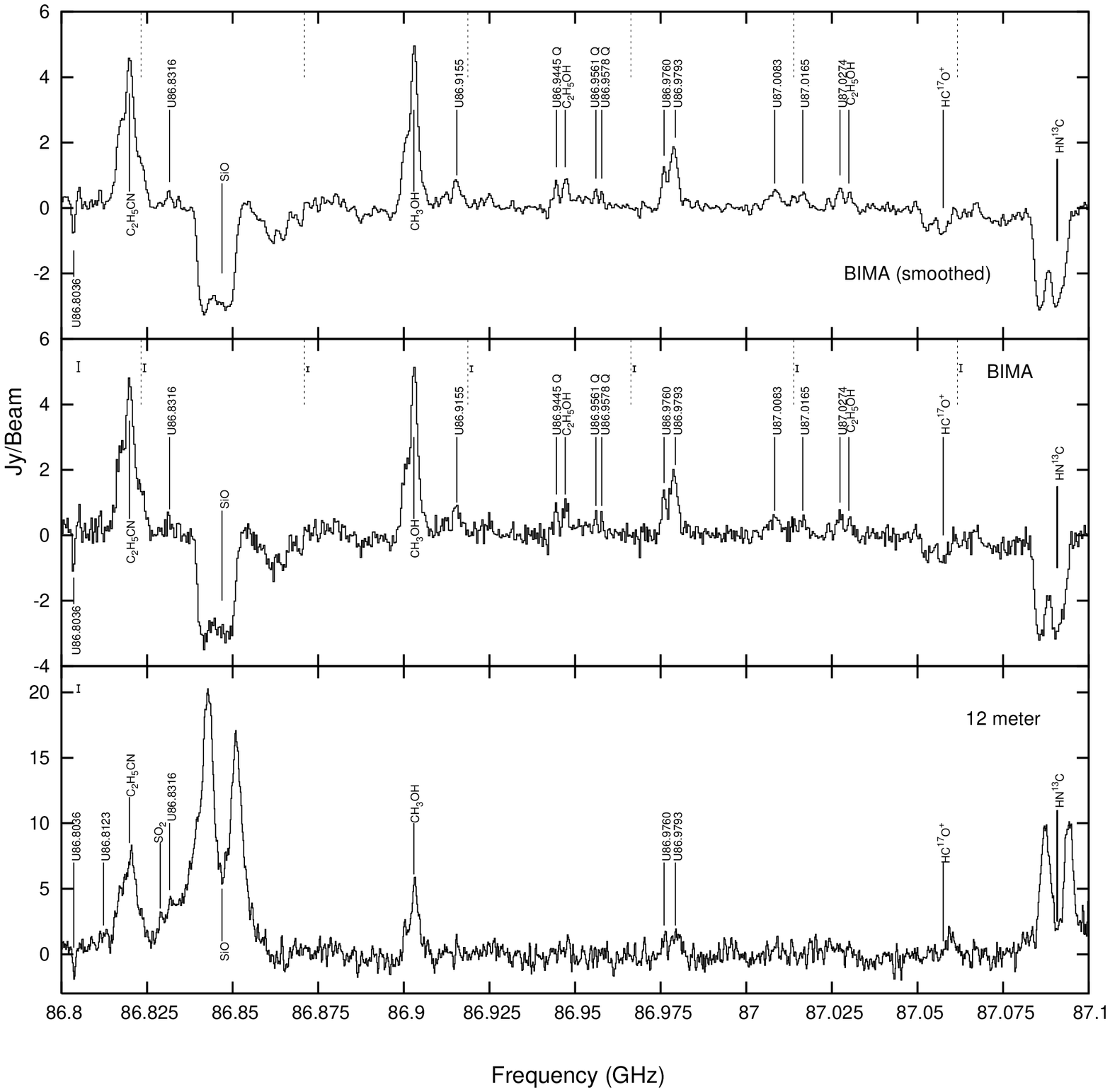}
\centerline{Figure \ref{fig:spec}d}
\end{figure}
\clearpage
\begin{figure}
\epsscale{1.1}
\plotone{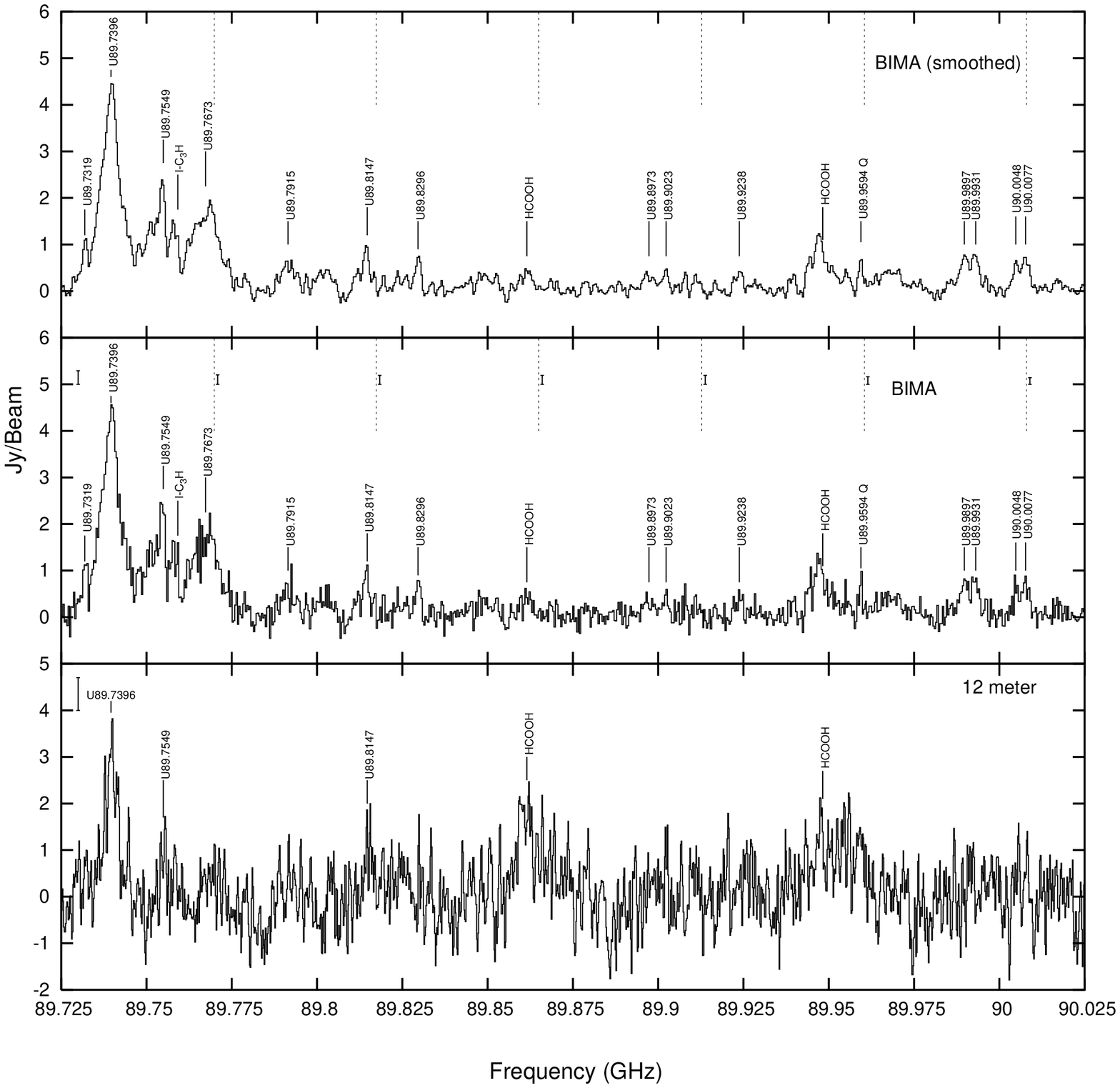}
\centerline{Figure \ref{fig:spec}e}
\end{figure}
\clearpage
\begin{figure}
\epsscale{1.1}
\plotone{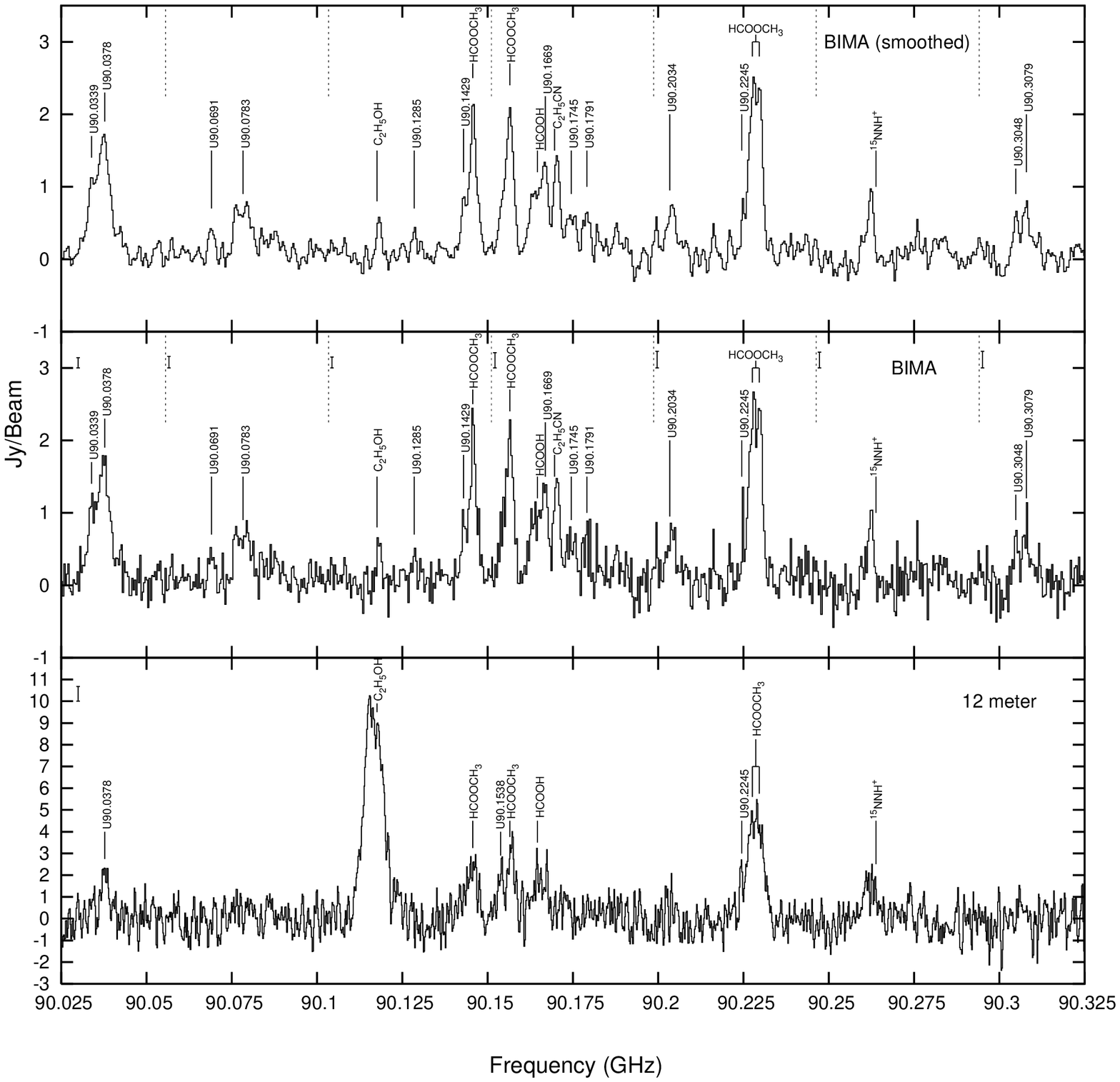}
\centerline{Figure \ref{fig:spec}f}
\end{figure}
\clearpage
\begin{figure}
\epsscale{1.1}
\plotone{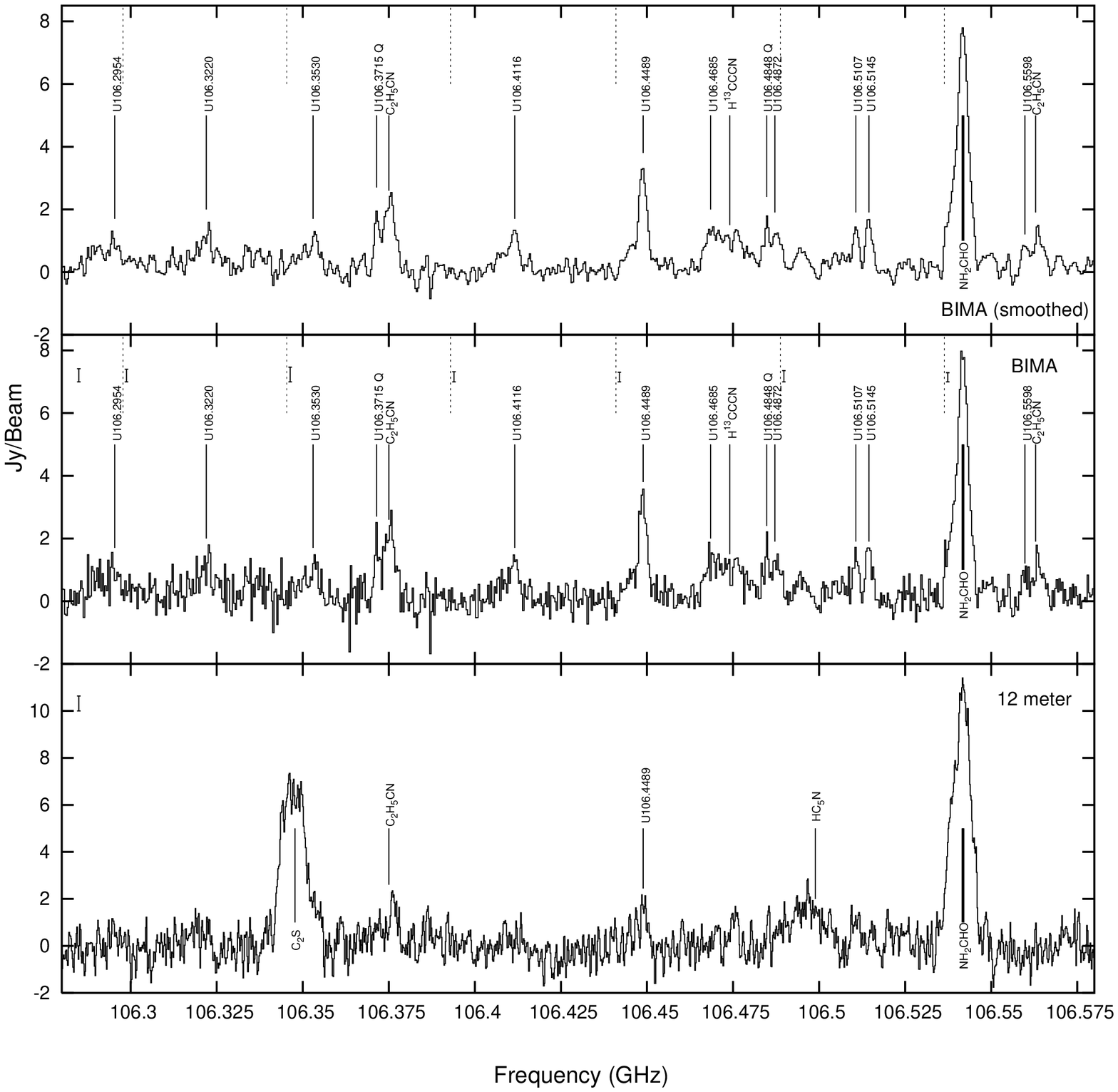}
\centerline{Figure \ref{fig:spec}g}
\end{figure}
\clearpage
\begin{figure}
\epsscale{1.1}
\plotone{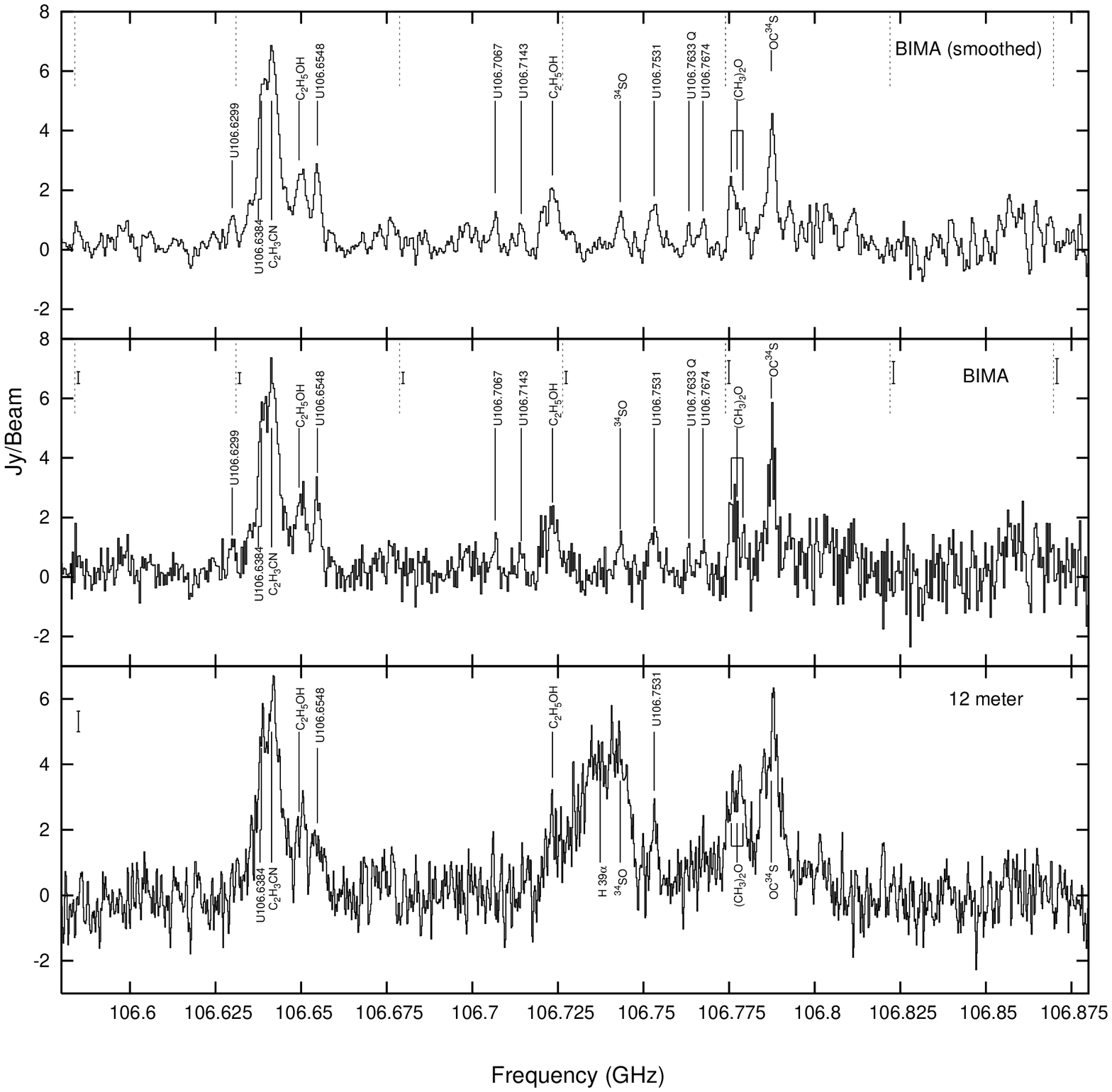}
\centerline{Figure \ref{fig:spec}h}
\end{figure}
\clearpage
\begin{figure}
\epsscale{1.1}
\plotone{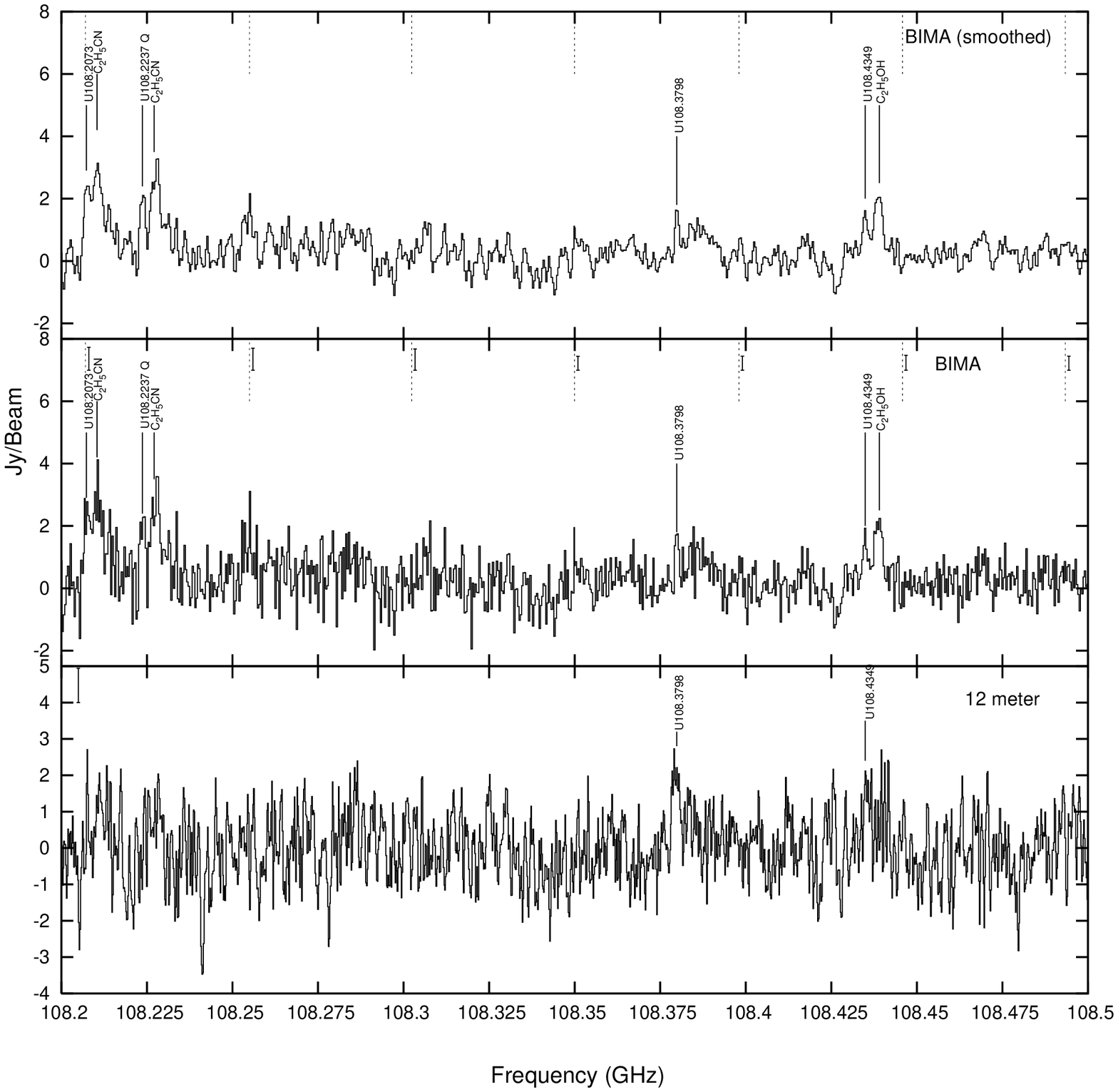}
\centerline{Figure \ref{fig:spec}i}
\end{figure}
\clearpage
\begin{figure}
\epsscale{1.1}
\plotone{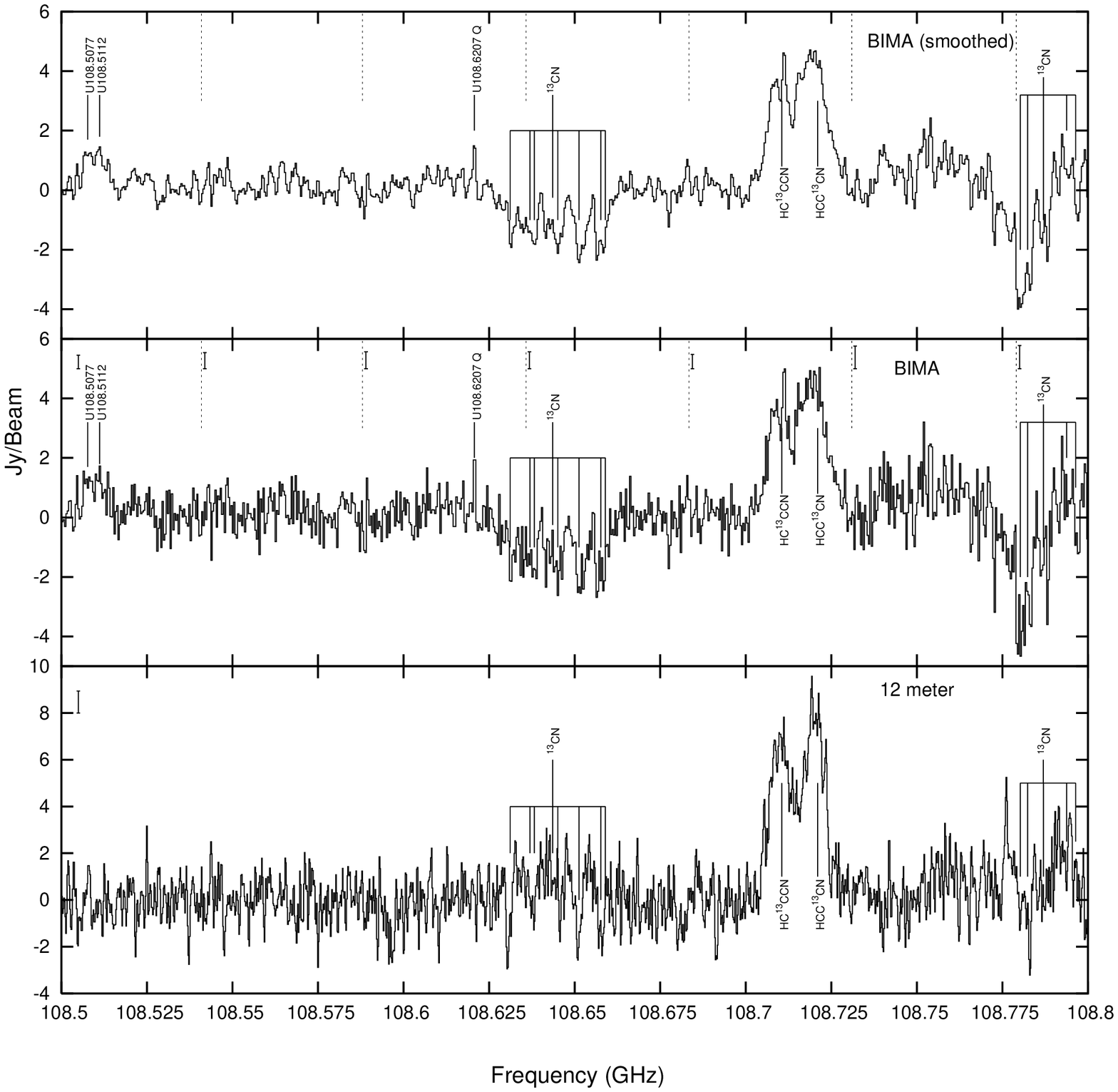}
\centerline{Figure \ref{fig:spec}j}
\end{figure}
\clearpage
\begin{figure}
\epsscale{1.1}
\plotone{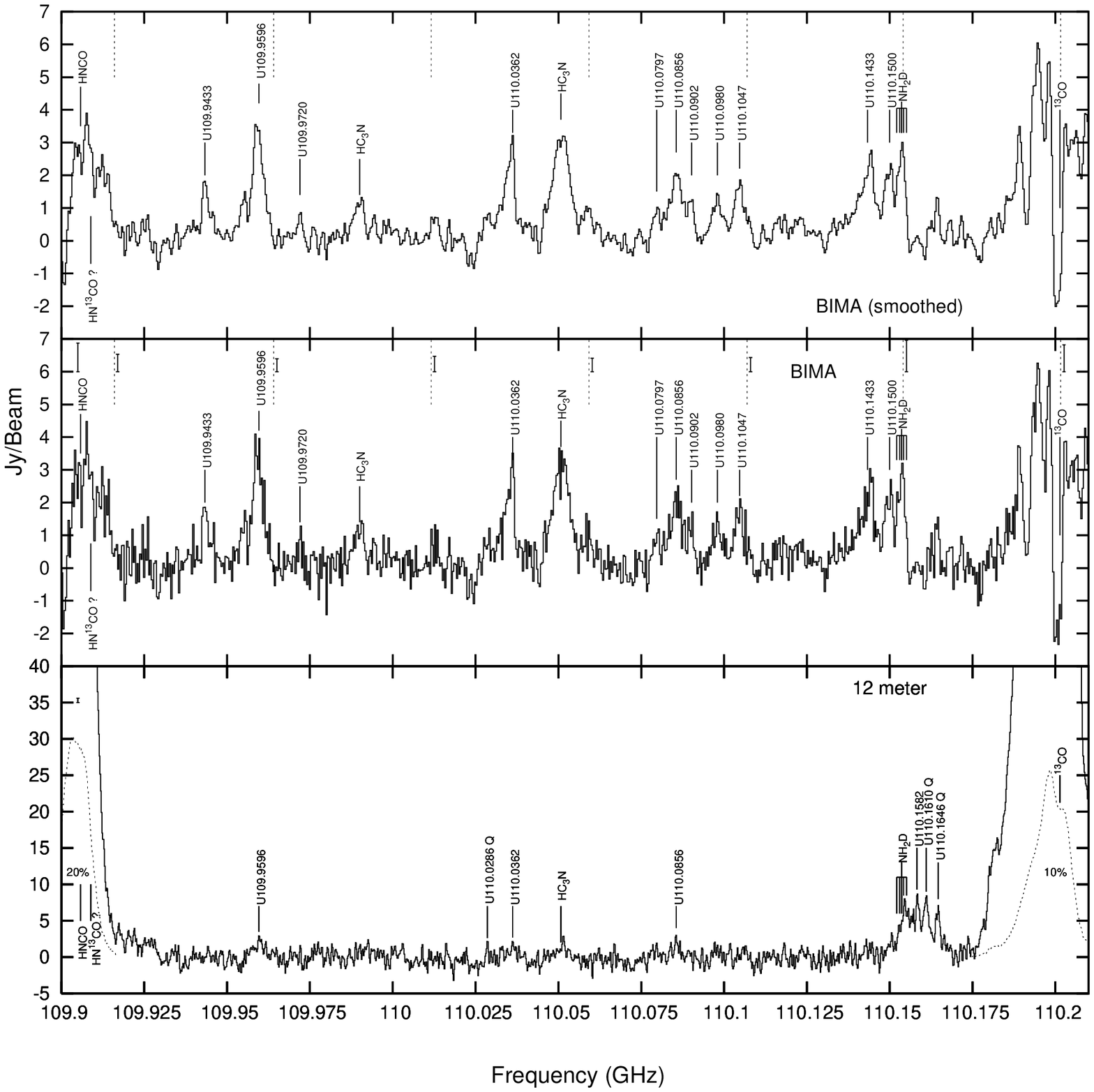}
\centerline{Figure \ref{fig:spec}k}
\end{figure}
\clearpage
\begin{figure}
\epsscale{1.1}
\plotone{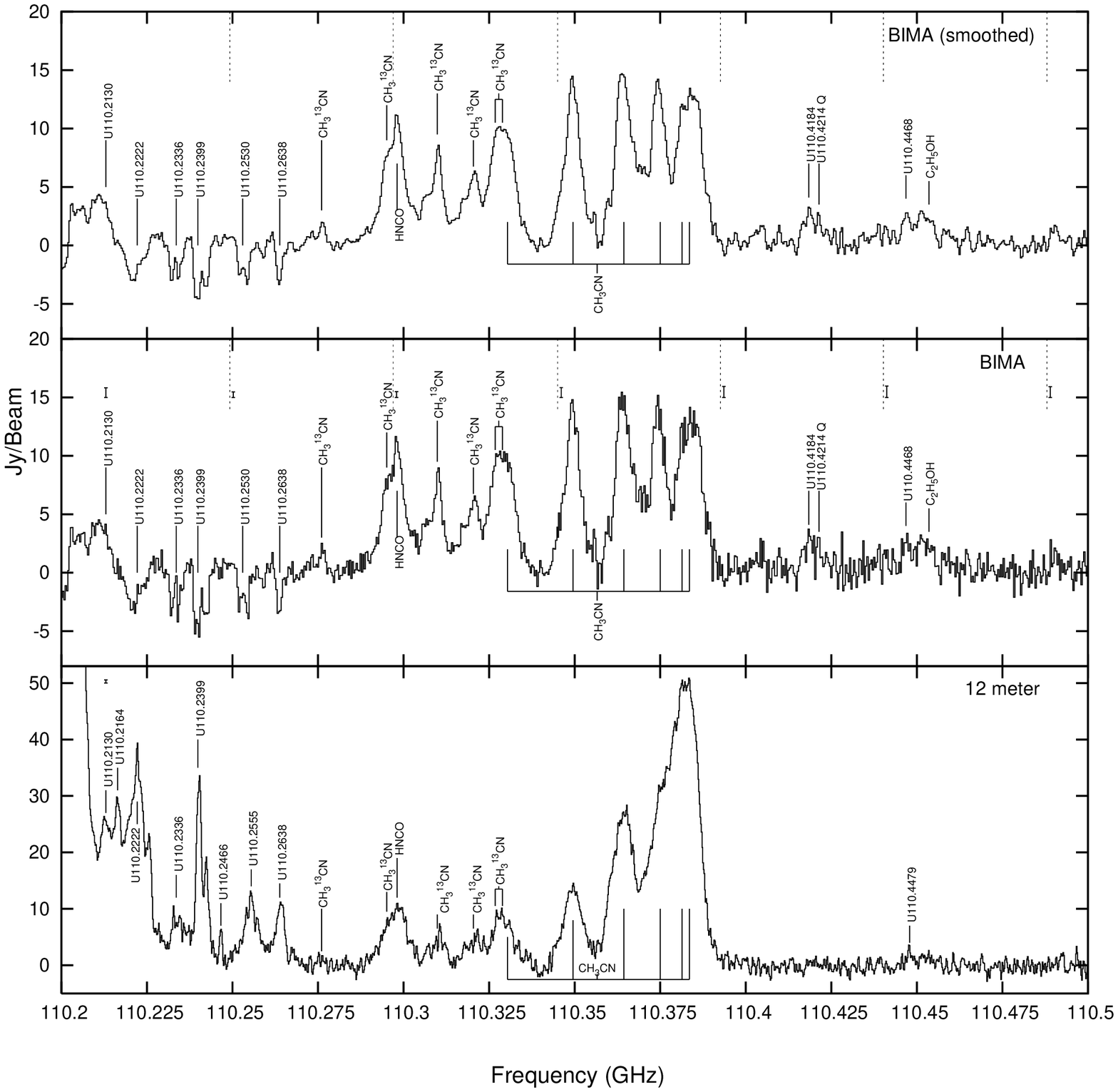}
\centerline{Figure \ref{fig:spec}l}
\end{figure}
\clearpage
\begin{figure}
\epsscale{0.9}
\plotone{f4.eps}
\centerline{Figure \ref{fig:mapa}}
\end{figure}

\clearpage
\begin{deluxetable}{lcc}
\tabletypesize{\scriptsize}
\tablewidth{0pt}
\tablecaption{Coordinates of Sources}
\tablehead{\colhead{Source} & \colhead{$\alpha$(J2000)} & \colhead{$\delta$(J2000)}}
\startdata
Sgr~B2(N-LMH)\tablenotemark{a} & $17^{h}47^{m}20^{s}.0$ & $-28{\degr}22{\arcmin}17{\arcsec}.3$\\
Sgr~B2(M)\tablenotemark{b} & $17^{h}47^{m}20^{s}.3$ & $-28{\degr}23{\arcmin}07{\arcsec}.3$\\
Sgr~B2(OH)\tablenotemark{b} & $17^{h}47^{m}20^{s}.8$ & $-28{\degr}23{\arcmin}32{\arcsec}.2$\\
Sgr~B2(S)\tablenotemark{b} & $17^{h}47^{m}20^{s}.3$ & $-28{\degr}23{\arcmin}46{\arcsec}.3$\\
Sgr~B2(NW)\tablenotemark{c} & $17^{h}47^{m}16^{s}.4$ & $-28{\degr}22{\arcmin}22{\arcsec}.5$\\
\enddata
\tablenotetext{a}{Taken from \citet{miao97}}
\tablenotetext{b}{Taken from \citet{sutton91}}
\tablenotetext{c}{Taken from \citet{nummelin98}}
\label{tab:sgrcoords}
\end{deluxetable}

\begin{deluxetable}{cccccc}
\tabletypesize{\scriptsize}
\tablecolumns{6}
\tablewidth{0pt}
\tablecaption{Beam Sizes\tablenotemark{a}\ \ and RMS Noise Levels of Observations}
\tablehead{
\colhead{} & \multicolumn{2}{c}{BIMA Array} & \colhead{} & \multicolumn{2}{c}{12 meter telescope}\\
\cline{2-3} \cline{5-6}\\
\colhead{Frequency} & \colhead{Beam\tablenotemark{b}} & \colhead{RMS\tablenotemark{b}} & \colhead{} &\colhead{Beam} & \colhead{RMS}\\
\colhead{(GHz.)} & \colhead{($\arcsec{\times}\arcsec$)} & \colhead{(Jy/beam)} & \colhead{} & \colhead{($\arcsec$)} & \colhead{(Jy/beam)}}
\startdata
86.200 & $27.2{\times}6.6$ & 0.216 & \ & 74 & 0.613\\
86.800 & $32.2{\times}6.5$ & 0.231 & \ & 73 & 0.569\\
90.025 & $25.2{\times}6.3$ & 0.197 & \ & 71 & 0.700\\
106.580 & $23.2{\times}5.2$ & 0.462 & \ & 60 & 0.628\\
108.500 & $21.0{\times}5.3$ & 0.586 & \ & 59 & 0.941\\
110.200 & $23.2{\times}5.1$ & 0.679 & \ & 57 & 0.598\\
\enddata
\tablenotetext{a}{Synthesized beam for BIMA Array observations and main diffraction beamwidth (FWHM) for 12 meter telescope observations.}
\tablenotetext{b}{Average of all scans.}
\label{tab:rms}
\end{deluxetable}

\begin{deluxetable}{lccc}
\tabletypesize{\scriptsize}
\tablecolumns{4}
\tablewidth{0pt}
\tablecaption{Summary of Molecules Detected by the BIMA Array and NRAO 12 Meter Radio Telescope}
\tablehead{
\colhead{} & \multicolumn{2}{c}{Detected Transitions} & \colhead{}\\
\cline{2-3} \colhead{} & \colhead{} &\colhead{} & \colhead {}\\
\colhead{Species} & \colhead{BIMA Array} & \colhead{12 meter telescope} & \colhead{Reference}}
\startdata
l-C$_3$H\tablenotemark{a} & 1 & 0 & 1,2,3\\
C$_2$H$_3$CN & 1 & 1 & 4\\
C$_2$H$_5$CN & 7 & 2 & 5\\
C$_2$H$_5$OH & 11 & 4 & 5,6\\
C$_2$S & 0 & 2 & 7,8\\
CH$_3$CN & 6 & 6 & 9\\
CH$_3^{13}$CN & 6 & 6 & 9\\
(CH$_3$)$_2$O & 6 & 4 & 10\\
CH$_3$OH & 2 & 2 & 11\\
$^{13}$CN\tablenotemark{a} & 4 & 4 & 12\\
$^{13}$CO & 1 & 1 & 13\\
H$^{39}\alpha$ & 0 & 1 & 14\\
HC$_5$N & 0 & 1 & 3,15\\
HC$_3$N & 2 & 1 & 15,16,17\\
H$^{13}$CCCN & 1 & 0 & 15,17\\
HC$^{13}$CCN & 1 & 1 & 16\\
HCC$^{13}$CN & 1 & 1 & 16\\
H$^{13}$CN\tablenotemark{a} & 1 & 1 & 18\\
HC$^{15}$N & 1 & 1 & 18\\
HCO\tablenotemark{a} & 0 & 1 & 3,19\\
H$^{13}$CO$^+$ & 1 & 1 & 20\\
HC$^{17}$O$^+$ & 1 & 1 & 21\\
HCOOCH$_3$ & 18 & 14 & 22\\
HCOOH & 4 & 4 & 23\\
HN$^{13}$C\tablenotemark{a} & 1 & 1 & 3,24\\
HNCO & 2 & 2 & 3,25\\
HN$^{13}$CO & 1 & 1 & 26\\
NH$_2$CHO\tablenotemark{a} & 2 & 2 & 26\\
NH$_2$D\tablenotemark{a} & 2 & 2 & 3,27,28\\
$^{15}$NNH$^+$ & 1 & 1 & 3,15\\
OC$^{34}$S & 1 & 1 & 3,18\\
SiO & 1 & 1 & 29\\
SO & 1 & 1 & 30\\
$^{34}$SO & 1 & 1 & 30\\
SO$_2$ & 2 & 2 & 31\\
U & 108 & 41 & \nodata\\
\hline
Totals & 199 & 116 & \\
\enddata
\tablenotetext{a}{Number of detected lines does not include hyperfine components.}
\tablerefs{(1) \citet{gott86} (2) \citet{yamamoto90b}; (3) \citet{jpl}; (4) \citet{gerry79}; (5) \citet{lovas82}; (6) J.~C.~Pearson, private communication; (7) \citet{yamamoto90a}; (8) \citet{murakami90}; (9) \citet{bou80}; (10) \citet{gron98}; (11) \citet{xu97}; (12) \citet{bog84}; (13) \citet{lovas74}; (14) \citet{lilley68}; (15) \citet{lovas03}; (16) \citet{laff78}; (17) \citet{wyr99}; (18) \citet{lovas78}; (19) \citet{snyder85}; (20) \citet{wood81}; (21) \citet{guel82}; (22) \citet{oest99}; (23) \citet{will80}; (24) \citet{frer79}; (25) \citet{winn76}; (26) \citet{john72}; (27) \citet{bester83}; (28) \citet{shah01}; (29) \citet{lovas74a}; (30) \citet{tiem74}; (31) \citet{lovas85}}
\label{tab:summ}
\end{deluxetable}

\begin{deluxetable}{llccccccc}
\tabletypesize{\tiny}
\tablecolumns{9}
\tablewidth{450pt}
\tablecaption{Detected Transitions}
\tablehead{
\colhead{Rest} & \colhead{} & \multicolumn{2}{c}{BIMA Array} & \colhead{} & \multicolumn{2}{c}{12 meter telescope} & \colhead{} & \colhead{}\\
\cline{3-4} \cline{6-7} \colhead{Frequency} & \colhead{} & \colhead{I$_0$} & \colhead{$\Delta$v} & \colhead{} & \colhead{I$_0$} & \colhead{$\Delta$v} & \colhead{$E_u$} & \colhead{$\langle$S$_{i,j}\mu^2\rangle$}\\
\colhead{(MHz)\tablenotemark{a}} & \colhead{Transition\tablenotemark{b}} & \colhead{(Jy bm$^{-1}$)} & \colhead{(km s$^{-1}$)} & \colhead{} & \colhead{(Jy bm$^{-1}$)} & \colhead{(km s$^{-1}$)} & \colhead{(K)} & \colhead{(D$^2$)}}
\startdata
\cutinhead{l-C$_3$H ($\nu_4 ^2{\Sigma}^{\mu}$)}
89,759.3280 (480)\tablenotemark{c} & $N_J=4_{\frac{7}{2}}-3_{\frac{5}{2}}, F=4-3$ & 1.0(8) & 7.0(56) & & $<0.7$ & \nodata & 49.9 & 18.45 \\
89,759.8490 (480)\tablenotemark{c} & $\;\;\;\;\;\;\;\;\;\;\;\;\;\;\;\;\;\;\;\;\;\;\;\;\,F=3-2$ & \nodata & \nodata & & $<0.7$ & \nodata & 49.9 & 13.74 \\
\cutinhead{C$_2$H$_3$CN (VyCN)}
106,641.3940 (170) & $11_{1,10}-10_{1,9}$ & 6.7(4) & 11.7(38) & & 6.1(4) & 10.7(18) & 32.9 & 147.71 \\
\cutinhead{C$_2$H$_5$CN (EtCN)}
86,484.2300 (400)\tablenotemark{d} & $29_{3,27}-28_{4,24}$ & 0.5(2) & 5.9(38) & & $<0.6$ & \nodata & 197.1 & 6.20 \\
86,819.8510 (130)\tablenotemark{e} & $10_{1,10}-9_{1,9}$ & 4.6(2) & 11.4(10) & & 7.5(2) & 13.8(14) & 24.1 & 146.68 \\
90,169.6330 (260)\tablenotemark{d} & $20_{2,18}-19_{3,17}$ & 1.5(2) & 6.6(18) & & $<0.7$ & \nodata & 96.4 & 6.09 \\
106,375.0030 (200) & $15_{3,12}-15_{2,13}$ & 2.4(4) & 10.9(26) & & 2.0(4) & 7.6(28) & 61.7 & 13.24 \\
106,562.9160 (320)\tablenotemark{d} & $27_{3,24}-26_{4,23}$ & 1.3(4) & 9.6(28) & & $<0.6$ & \nodata & 174.1 & 7.23 \\
108,210.3860 (170)\tablenotemark{d} & $11_{1,11}-10_{0,10}$ & 2.7(6) & 13.8(70) & & $<0.9$ & \nodata & 28.6 & 11.01 \\*
108,227.0930 (200)\tablenotemark{d} & $14_{3,11}-14_{2,12}$ & 2.6(8) & 13.0(52) & & $<0.9$ & \nodata & 55.2 & 12.07 \\*
\cutinhead{C$_2$H$_5$OH\tablenotemark{f} (EtOH)}
86,311.2673 (32)\tablenotemark{d} & $5_{2,4}^+-4_{2,3}^+$ &$<0.3$ &\nodata && 1.8(2) & 6.0(10) & 73.7 & 5.03 \\
86,555.9124 (28)\tablenotemark{c,d} & $5_{4,2}^+-4_{4,1}^+$ &0.7(4) &4.4(26) && $<0.6$ &\nodata & 88.3 & 4.20 \\
86,556.0115 (28)\tablenotemark{c,d} & $5_{4,1}^+-4_{4,0}^+$ &\nodata &\nodata &&\nodata &\nodata & 88.3 & 4.20 \\
86,604.2917 (28)\tablenotemark{d} & $5_{3,3}^+-4_{3,2}^+$ &0.9(2) &4.3(16) &&$<0.6$ &\nodata & 79.8 & 4.07 \\
86,621.7341 (28)\tablenotemark{d} & $5_{3,2}^+-4_{3,1}^+$ &0.9(2) &7.5(24) &&$<0.6$ &\nodata & 79.8 & 4.07 \\
86,947.1200 (270)\tablenotemark{d} & $15_{2,13}-15_{1,14}$ & 0.9(4) & 7.2(30) & & $<0.6$ & \nodata & 108.1 & 22.56 \\
87,029.9919 (30)\tablenotemark{d} & $5_{2,3}^+-4_{2,2}^+$ & 0.6(2)&4.2(22) &&$<0.6$ &\nodata & 73.8 & 5.02 \\
90,117.6000 (90) & $4_{1,4}-3_{0,3}$ & 0.6(2) & 4.7(22) & & 9.9(4) & 21.5(10) & 9.4 & 5.35 \\
106,649.4400 (310)\tablenotemark{d} & $13_{1,12}-13_{0,13}$ & 2.7(4) & 11.4(32) & & 2.5(4) & 7.8(26) & 79.4 & 10.14 \\
106,723.4100 (180) & $9_{2,8}-9_{1,9}$ & 1.9(4) & 11.3(44) & & 1.7(4) & 10.4(34) & 42.7 & 7.66 \\*
108,439.0700 (800)\tablenotemark{d} & $13_{3,10}-13_{2,11}$ & 2.2(2) & 7.8(12) & & $<0.6$ & \nodata & 88.2 & 18.19 \\*
110,453.5300 (3990)\tablenotemark{d} & $21_{3,18}-21_{2,19}$ & 2.7(6) & 15.4(48) & & $<0.6$ & \nodata & 208.7 & 33.93 \\*
\cutinhead{C$_2$S}
86,181.4130 (100) & $J_N=6_7-5_6$ & $<0.2$ & \nodata & & 2.8(4) & 18.7(24) & 23.3 & 41.03 \\*
106,347.7420 (100) & $J_N=9_8-8_7$ & $<0.5$ & \nodata & & 7.1(4) & 22.4(12) & 24.5 & 63.01 \\*
\cutinhead{CH$_3$CN}
110,330.3470 (23)\tablenotemark{g} & $J_K=6_{5}-5_{5}$ & 6.0(12) & 19.8(10) & & 3.1(10) & 20.9(6) & 196.1 & 28.07 \\
110,349.4730 (16) & $J_K=6_{4}-5_{4}$ & 13.1(10) & 15.8(14) & & 12.5(12) & 20.9(6) & 132.2 & 51.03 \\
110,364.3570 (11) & $J_K=6_{3}-5_{3}$ & 14.1(10) & 17.7(16) & & 26.8(12) & 20.9(6) & 82.5 & 68.90 \\
110,374.9920 (9) & $J_K=6_{2}-5_{2}$ & 12.6(10) & 18.0(22) & & 27.1(18) & 20.9(6) & 47.0 & 81.66 \\*
110,381.3760 (10) & $J_K=6_{1}-5_{1}$ & 10.7(18) & 12.0(16) & & 16.1(42) & 20.9(6) & 25.7 & 89.31 \\*
110,383.5040 (10) & $J_K=6_{0}-5_{0}$ & 10.9(18) & 12.0(16) & & 36.5(42) & 20.9(6) & 18.5 & 91.87 \\*
\cutinhead{CH$_3^{13}$CN}
110,276.0470 (460)\tablenotemark{d} & $J_K=6_{5}-5_{5}$ & 1.2(4) & 17.2(4) & & 0.6(4) & 15.4(8) & 197.5 & 28.07 \\
110,295.0700 (340)\tablenotemark{d} & $J_K=6_{4}-5_{4}$ & 3.2(18) & 17.2(4) & & 5.3(4) & 15.4(8) & 133.1 & 51.03 \\
110,309.8670 (290) & $J_K=6_{3}-5_{3}$ & 6.8(4) & 17.2(4) & & 4.5(4) & 15.4(8) & 83.0 & 68.90 \\
110,320.4350 (290)\tablenotemark{d} & $J_K=6_{2}-5_{2}$ & 5.4(4) & 17.2(4) & & 4.4(4) & 15.4(8) & 47.2 & 81.66 \\*
110,326.7770 (300)\tablenotemark{g} & $J_K=6_{1}-5_{1}$ & 1.6(12) & 17.2(4) & & 1.5(8) & 15.4(8) & 25.7 & 89.31 \\*
110,328.8900 (310)\tablenotemark{g} & $J_K=6_{0}-5_{0}$ & 4.3(22) & 17.2(4) & & 4.9(16) & 15.4(8) & 18.5 & 91.87 \\*
\cutinhead{(CH$_3$)$_2$O}
85,973.2490 (80) & $13_{2,12}-12_{3,9}$AA & 0.3(2) & 5.6(30) & & $<0.6$ & \nodata & 88.0 & 2.81 \\
85,976.1310 (80) & $13_{2,12}-12_{3,9}$EE & 0.8(4) & 5.6(30) & & $<0.6$ & \nodata & 88.0 & 2.81 \\
106,775.6790 (80) & $9_{1,8}-8_{2,7}$AA & 0.9\tablenotemark{h} & 7.0 & & 1.1(2) & 15.7(26) & 43.4 & 3.66 \\
106,777.3710 (60) & $9_{1,8}-8_{2,7}$EE & 1.4\tablenotemark{h} & 7.0 & & 1.7(2) & 15.7(26) & 43.4 & 3.66 \\*
106,779.0580 (60) & $9_{1,8}-8_{2,7}$AE & 0.5\tablenotemark{h} & 7.0 & & 0.6(1) & 15.7(26) & 43.4 & 3.66 \\*
106,779.0660 (60) & $9_{1,8}-8_{2,7}$EA & 0.4\tablenotemark{h} & 7.0 & & 0.4(1) & 15.7(26) & 43.4 & 3.66 \\*
\cutinhead{CH$_3$OH}
86,615.6020 (140)\tablenotemark{e} & $7_{2,6}-6_{3,3}$A$-$ & 6.3(4) & 10.2(8) & & 5.8(4) & 8.3(10) & 102.7 & 1.36 \\*
86,902.9470 (140)\tablenotemark{e} & $7_{2,5}-6_{3,4}$A$+$ & 5.0(2) & 8.9(6) & & 5.2(4) & 8.9(8) & 102.7 & 1.36 \\*
\cutinhead{$^{13}$CN}
108,631.1210 (1000)\tablenotemark{d} & $N_{J,F}=1_{1,0}-0_{0,1}$ & \tablenotemark{i} & \nodata & & \tablenotemark{i} & \nodata & 5.2 & 2.10 \\
108,636.9230 (1000)\tablenotemark{d} & $N_{J,F}=1_{1,1}-1_{0,1}$ & \tablenotemark{i} & \nodata & & \tablenotemark{i} & \nodata & 5.2 & 2.10 \\
108,638.2120 (1000)\tablenotemark{d} & $N_{J,F}=1_{1,1}-0_{1,0}$ & \tablenotemark{i} & \nodata & & \tablenotemark{i} & \nodata & 5.2 & 2.10 \\
108,643.5900 (1000)\tablenotemark{d} & $N_{J,F}=1_{1,2}-0_{1,1}$ & \tablenotemark{i} & \nodata & & \tablenotemark{i} & \nodata & 5.2 & 2.10 \\
108,645.0640 (2000)\tablenotemark{d} & $N_{J,F}=1_{1,1}-0_{1,1}$ & \tablenotemark{i} & \nodata & & \tablenotemark{i} & \nodata & 5.2 & 2.10 \\
108,645.0640 (2000)\tablenotemark{d} & $N_{J,F}=1_{1,0}-0_{1,1}$ & \tablenotemark{i} & \nodata & & \tablenotemark{i} & \nodata & 5.2 & 2.10 \\
108,651.2970 (1000) & $N_{J,F}=1_{1,2}-0_{0,1}$ & \tablenotemark{i} & \nodata & & \tablenotemark{i} & \nodata & 5.2 & 2.10 \\
108,657.6460 (1000) & $N_{J,F}=1_{1,2}-0_{1,2}$ & \tablenotemark{i} & \nodata & & \tablenotemark{i} & \nodata & 5.2 & 2.10 \\
108,658.9480 (1000) & $N_{J,F}=1_{1,1}-0_{1,2}$ & \tablenotemark{i} & \nodata & & \tablenotemark{i} & \nodata & 5.2 & 2.10 \\
108,780.2010 (1000) & $N_{J,F}=1_{2,3}-0_{1,2}$ & \tablenotemark{i} & \nodata & & \tablenotemark{i} & \nodata & 5.2 & 4.21 \\
108,782.3740 (1000) & $N_{J,F}=1_{2,2}-0_{1,1}$ & \tablenotemark{i} & \nodata & & \tablenotemark{i} & \nodata & 5.2 & 4.21 \\
108,786.9820 (1000) & $N_{J,F}=1_{2,1}-0_{1,0}$ & \tablenotemark{i} & \nodata & & \tablenotemark{i} & \nodata & 5.2 & 4.21 \\
108,793.7530 (1000)\tablenotemark{d} & $N_{J,F}=1_{2,1}-0_{1,1}$ & \tablenotemark{i} & \nodata & & \tablenotemark{i} & \nodata & 5.3 & 4.21 \\
108,796.4000 (1000)\tablenotemark{d} & $N_{J,F}=1_{2,2}-0_{1,2}$ & \tablenotemark{i} & \nodata & & \tablenotemark{i} & \nodata & 5.3 & 4.21 \\*
\cutinhead{$^{13}$CO}
110,201.3700 (200) & $J=1-0$ & \tablenotemark{i} & \nodata & & 123.5(132)\tablenotemark{j} & 12.1(10) & 5.3 & 0.01 \\
 & & & & & 105.8(122)\tablenotemark{j} & 10.8(6) & & \\
 & & & & & 156.2(64)\tablenotemark{j} & 34.5(16) & & \\
\cutinhead{H$^{39}\alpha$}
106,737.3630 & Recombination Line & $<0.4$ & \nodata & & 4.0(2) & 43.2(54) & \nodata & \nodata \\
\cutinhead{HC$_5$N}
106,498.9110 (80) & $J=40-39$ & $<0.4$ & \nodata & & 1.7(2) & 34.1(52) & 103.6 & 749.96 \\
\cutinhead{HC$_3$N}
109,990.0000 (2000) & $J=12-11, \nu^1_6, \nu^1_7, \ell=1e-$ & 1.3(2)&10.0(26)&& $<0.6$&\nodata & 1391.0 & 166.00\tablenotemark{k}\\*
110,050.7650 (184) & $J=12-11, 3\nu^1_7, \ell=1e$ & 3.1(4) & 18.2(26) & & 1.6(6) & 8.3(40) & 990.7 & 165.27 \\*
\cutinhead{H$^{13}$CCCN}
106,474.1000 (10000) & $J=12-11, \nu^2_7$ & 1.1(2) & 33.0(116) & & $<0.6$ &\nodata & 673.0 & 166.00\tablenotemark{k}\\
\cutinhead{HC$^{13}$CCN}
108,710.5230 (240) & $J=12-11$ & 3.5(2) & 17.9(32) & & 6.9(4) & 20.3(18) & 33.9 & 166.42 \\
\cutinhead{HCC$^{13}$CN}
108,721.0077 (144) & $J=12-11$ & 4.6(4) & 27.7(36) & & 8.1(4) & 19.2(14) & 33.9 & 166.42 \\
\cutinhead{H$^{13}$CN}
86,338.7670 (600) & $J=1-0\ F=1-1$ & \tablenotemark{i} & \nodata & & \tablenotemark{i} & \nodata & 4.0 & 8.91 \\*
86,340.1840 (600) & $J=1-0\ F=2-1$ & \tablenotemark{i} & \nodata & & \tablenotemark{i} & \nodata & 4.0 & 8.91 \\*
86,342.2740 (600) & $J=1-0\ F=0-1$ & \tablenotemark{i} & \nodata & & \tablenotemark{i} & \nodata & 4.0 & 8.91 \\*
\cutinhead{HC$^{15}$N}
86,054.9610 (600) & $J=1-0$ & \tablenotemark{i} & \nodata & & \tablenotemark{i} & \nodata & 4.1 & 8.91 \\
\cutinhead{HCO}
 & $N_{K_a,K_c}=1_{0,1}-0_{0,0},J=\frac{3}{2}-\frac{1}{2}$ & & & & & & & \\
86,670.8200 (800) & $F=2-1$ & $<0.3$ & \nodata & & 2.8(4) & 19.7(26) & 4.2 & 1.86 \\*
86,708.3500 (800) & $F=1-0$ & $<0.3$ & \nodata & & 2.1(2) & 17.8(28) & 4.2 & 1.86 \\*
\cutinhead{H$^{13}$CO$^{+}$}
86,754.3290 (780) & $J=1-0$ & \tablenotemark{i} & \nodata & & \tablenotemark{i} & \nodata & 4.2 & 10.89 \\
\cutinhead{HC$^{17}$O$^+$}
87,057.5000 (10000) & $J=1-0$ & \tablenotemark{i} & \nodata & & \tablenotemark{i} & \nodata & 4.2 & 10.89 \\
\cutinhead{HCOOCH$_3$ (MeF)}
85,919.0860 (280) & $7_{6,1}-6_{6,0}$E & 0.7(3)\tablenotemark{l} & 6.8(23) & & $<0.6$ & \nodata & 40.4 & 5.00 \\
85,926.5080 (220) & $7_{6,2}-6_{6,1}$E & 0.7(3)\tablenotemark{l} & 6.8(23) & & $<0.6$ & \nodata & 40.4 & 5.00 \\
85,927.2300 (240) & $7_{6,2}-6_{6,1}$A & 0.7(3)\tablenotemark{l} & 6.8(23) & & $<0.6$ & \nodata & 40.4 & 5.00 \\
85,927.2360 (240) & $7_{6,1}-6_{6,0}$A & 0.7(3)\tablenotemark{l} & 6.8(23) & & $<0.6$ & \nodata & 40.4 & 5.00 \\
86,021.0080 (260) & $7_{5,2}-6_{5,1}$E & 1.2(4) & 5.6(22) & & 2.6(8) & 3.6(12) & 33.1 & 9.20 \\
86,027.6740 (220) & $7_{5,3}-6_{5,2}$E & 1.5(4) & 7.8(40) & & 1.2(6) & 5.9(20) & 33.1 & 9.20 \\
86,029.4450 (240) & $7_{5,3}-6_{5,2}$A & 1.2(4) & 4.2(28) & & 1.2(6) & 5.9(20) & 33.1 & 9.20 \\
86,030.2120 (240) & $7_{5,2}-6_{5,1}$A & 1.8(6) & 4.4(18) & & 1.2(8) & 5.9(20) & 33.1 & 9.20 \\
86,210.0790 (240) & $7_{4,4}-6_{4,3}$A & 1.3(2) & 8.2(20) & & 1.7(4) & 8.3\tablenotemark{m} & 27.2 & 12.70 \\
86,223.5480 (260) & $7_{4,3}-6_{4,2}$E & 1.5(2) & 7.1(32) & & 1.2(6) & 8.3\tablenotemark{m} & 27.2 & 12.60 \\
86,224.1060 (220) & $7_{4,4}-6_{4,3}$E & 1.3(2) & 3.7(16) & & 1.7(4) & 8.3\tablenotemark{m} & 27.2 & 12.60 \\
86,250.5760 (240) & $7_{4,3}-6_{4,2}$A & 1.5(2) & 7.8(26) & & 3.1(4) & 14.4(20) & 27.2 & 12.70 \\
86,265.8260 (240) & $7_{3,5}-6_{3,4}$A & 1.7(4) & 6.3(24) & & 2.9(4) & 8.4(14) & 22.4 & 15.30 \\
86,268.6590 (220) & $7_{3,5}-6_{3,4}$E & 1.3(4) & 8.6(40) & & 2.5(4) & 5.2(40) & 22.6 & 15.20 \\
90,145.6340 (240) & $7_{2,5}-6_{2,4}$E & 2.0(2) & 9.1(18) & & 2.3(4) & 13.6(28) & 19.7 & 17.30 \\
90,156.5110 (260) & $7_{2,5}-6_{2,4}$A & 1.9(2) & 11.1(16) & & 3.3(6) & 8.2(20) & 19.7 & 17.30 \\
90,227.5950 (260) & $8_{0,8}-7_{0,7}$E & 2.2(4) & 7.7(10) & & 2.8(30) & 14.0(18) & 20.1 & 21.00 \\
90,229.6470 (280) & $8_{0,8}-7_{0,7}$A & 2.2(4) & 7.7(10) & & 3.0(26) & 14.0(18) & 20.0 & 21.00 \\*
\cutinhead{HCOOH}
86,546.1800 (200) & $4_{1,4}-3_{1,3}$ & 0.5(2) & 9.4(44)\tablenotemark{h} & & 2.4(4) & 10.8(44) & 13.6 & 7.26 \\
89,861.4800 (200) & $4_{2,3}-3_{2,2}$ & 0.4(2) & 14.1(48) & & 2.3(8) & 8.5(33)\tablenotemark{h} & 23.5 & 5.80 \\
89,948.2100 (200)\tablenotemark{d} & $4_{3,2}-3_{3,1}$ & 1.3(2) & 9.2(36) & & 1.8(2) & 7.2(14) & 39.4 & 3.39 \\*
90,164.6200 (200)\tablenotemark{d} & $4_{2,2}-3_{2,1}$ & 0.9(2) & 13.5(80) & & 1.9(4) & 17.8(42) & 23.5 & 5.80 \\*
\cutinhead{HN$^{13}$C}
87,090.7350 (920) & $J=1-0\ F=0-1$ & \tablenotemark{i} & \nodata & & \tablenotemark{i} & \nodata & 4.2 & 7.28 \\*
87,090.8590 (920) & $J=1-0\ F=2-1$ & \tablenotemark{i} & \nodata & & \tablenotemark{i} & \nodata & 4.2 & 7.28 \\*
87,090.9420 (920) & $J=1-0\ F=1-1$ & \tablenotemark{i} & \nodata & & \tablenotemark{i} & \nodata & 4.2 & 7.28 \\*
\cutinhead{HNCO}
109,905.7530 (100)\tablenotemark{j} & $5_{0,5}-4_{0,4}$ & 2.9(4) & 24.4(74) & & 151.2(12) & 26.5(2) & 15.8 & 12.40 \\*
 & & 1.7(12) & 6.3(56) & & 23.6(22) & 6.8(8) & & \\
110,298.0980 (80)\tablenotemark{h} & $5_{1,4}-4_{1,3}$ & 8.5(8) & 17.6(14) & & 8.4(6) & 15.0(10) & 59.0 & 11.91 \\*
\cutinhead{HN$^{13}$CO}
109908.8710 (440)\tablenotemark{d} & $5_{0,5}-4_{0,4}$ & ??? & \nodata & & ??? & \nodata & 15.8 & 12.40 \\*
\cutinhead{NH$_2$CHO}
86,381.9540 (80)\tablenotemark{d} & $7_{1,6}-7_{0,7}, F=7-7$ & 0.7\tablenotemark{h} & 7.5(40) & & 0.4\tablenotemark{h} & 7.2(56) & 32.5 & 1.46\tablenotemark{n} \\
86,383.2490 (30)\tablenotemark{c,d} & $7_{1,6}-7_{0,7}, F=8-8$ & 1.5(6) & 7.5(40) & & 1.6(6) & 7.2(56) & 32.5 & 1.69\tablenotemark{n} \\
86,383.4350 (50)\tablenotemark{c,d} & $7_{1,6}-7_{0,7}, F=6-6$ & \nodata & \nodata & & \nodata & \nodata & 32.5 & 1.28\tablenotemark{n} \\
106,541.8110 (800) & $5_{2,3}-4_{2,2}$\tablenotemark{e} & 7.4(4) & 12.5(8) & & 10.6(4) & 17.8(6) & 27.2 & 54.92 \\*
\cutinhead{NH$_2$D}
Ortho & $1_{0,1}-1_{1,1}$ & & & & & & & \\
85,924.7470 (400) & $F=0-1$ & 0.1(0) & 18.0(48) & & 0.3(0) & 26.0(28) & 20.7 & 0.36\tablenotemark{n} \\
85,925.6840 (400) & $F=2-1$ & 0.2(0) & 18.0(48) & & 0.4(0) & 26.0(28) & 20.7 & 0.45\tablenotemark{n} \\
85,926.2630 (200) & $F=0-0$ & 0.0(0) & 18.0(48) & & 0.0(0) & 26.0(28) & 20.7 & 0.00\tablenotemark{n} \\
85,926.2630 (200) & $F=1-1$ & 0.5(1) & 18.0(48) & & 1.2(1) & 26.0(28) & 20.7 & 1.34\tablenotemark{n} \\
85,926.2630 (200) & $F=2-2$ & 0.1(0) & 18.0(48) & & 0.2(0) & 26.0(28) & 20.7 & 0.27\tablenotemark{n} \\
85,926.8580 (400) & $F=1-2$ & 0.2(0) & 18.0(48) & & 0.4(0) & 26.0(28) & 20.7 & 0.45\tablenotemark{n} \\
85,927.7210 (400) & $F=1-0$ & 0.1(0) & 18.0(48) & & 0.3(0) & 26.0(28) & 20.7 & 0.36\tablenotemark{n} \\
Para & $1_{0,1}-1_{1,1}$ & & & & & & & \\
110,152.0840 (400) & $F=0-1$ & 0.5(2) & 4.4(20) & & 0.7(3) & 2.3(11) & 21.2 & 0.36\tablenotemark{n} \\
110,152.9950 (400) & $F=2-1$ & 0.6(2) & 4.4(20) & & 0.9(4) & 2.3(11) & 21.2 & 0.45\tablenotemark{n} \\
110,153.5990 (200) & $F=0-0$ & 0.0(0) & 4.4(20) & & 0.0(0) & 2.3(11) & 21.2 & 0.00\tablenotemark{n} \\
110,153.5990 (200) & $F=1-1$ & 1.8(6) & 4.4(20) & & 2.8(12) & 2.3(11) & 21.2 & 1.34\tablenotemark{n} \\
110,153.5990 (200) & $F=2-2$ & 0.3(1) & 4.4(20) & & 0.5(2) & 2.3(11) & 21.2 & 0.27\tablenotemark{n} \\*
110,154.2220 (400) & $F=1-2$ & 0.6(2) & 4.4(20) & & 0.9(4) & 2.3(11) & 21.2 & 0.45\tablenotemark{n} \\*
110,155.0530 (400) & $F=1-0$ & 0.5(2) & 4.4(20) & & 0.7(3) & 2.3(11) & 21.2 & 0.36\tablenotemark{n} \\*
\cutinhead{$^{15}$NNH$^{+}$}
90,263.8330 (600) & $J=1-0$ & 1.0(2) & 6.3(20) & & 1.9(2) & 14.2(28) & 4.5 & 11.56\\
\cutinhead{OC$^{34}$S}
106,787.3800 (1600) & $J=9-8$ & 3.8(10) & 10.1(28) & & 4.8(4) & 19.2(18) & 26.3 & 4.60 \\
\cutinhead{SiO}
86,846.8910 (560) & $J=2-1$ & \tablenotemark{i} & \nodata & & \tablenotemark{i} & \nodata & 6.3 & 19.20 \\
\cutinhead{SO}
86,093.9380 (340) & $J_K=2_{2}-1_{1}$\tablenotemark{e} & 7.6(4) & 14.3(10) & & 19.8(2) & 20.6(4) & 19.3 & 3.60 \\
\cutinhead{$^{34}$SO}
106,743.2440 (1400) & $J_N=2_{3}-1_{2}$ & 1.2(6) & 6.7(32) & & 2.0(8) & 14.3(54) & 20.9 & 3.63 \\
\cutinhead{SO$_2$}
86,153.7090 (250) & $39_{9,31}-40_{8,32}$ & 0.8(4) & 4.0(26) & & $<0.6$ & \nodata & 916.1 & 15.91 \\*
86,639.0400 (1000) & $8_{3,5}-9_{2,8}$ & 2.0(2) & 20.7(38) & & 1.5(4) & 13.0(34) & 55.2 & 3.02 \\*
86,828.8820 (590)\tablenotemark{d} & $20_{2,18}-21_{1,21}$ & $<0.2$ & \nodata & & 1.9\tablenotemark{h} & 6.8 & 207.8 & 0.07 \\*

\enddata
\tablenotetext{a}{Uncertainty is 95\% (2$\sigma$) confidence level.}
\tablenotetext{b}{Transitions are $J_{K_a,K_c}$ unless otherwise noted.}
\tablenotetext{c}{These transitions were blended and were fit with a single Gaussian.}
\tablenotetext{d}{Previously undetected transition.}
\tablenotetext{e}{Multiple emission components are visible. Only the intensity and line width of the main component are given.}
\tablenotetext{f}{ For EtOH a ``+''=gauche+ and a ``-''=gauche- state.}
\tablenotetext{g}{These transitions were highly blended and thus the intensity and line widths were approximated because the least squares Gaussian fitting did not give a satisfactory fit.}
\tablenotetext{h}{Intensity and line width were approximated because the least squares Gaussian fitting did not give a satisfactory fit.}
\tablenotetext{i}{Multiple emission and absorption components detected.}
\tablenotetext{j}{Multiple strong components are apparent in the data (see Figure 2), each component is listed on a separate line in this table.}
\tablenotetext{k}{Estimated value since the dipole moment $\mu$ was not stated in the reference.}
\tablenotetext{l}{Due to confusion from transitions from NH$_2$D the intensity and line width of this transition of MeF was extrapolated from the other detected MeF transitions.}
\tablenotetext{m}{Line width was fixed in order to get an adequate intensity fit.}
\tablenotetext{n}{In order to calculate the line strength of these hyperfine components the line strength for the entire transition was multiplied by the relative intensity of the hyperfine component.}
\label{tab:mol}
\end{deluxetable}

\begin{deluxetable}{ccccccl}
\tabletypesize{\tiny}
\tablecolumns{7}
\tablewidth{502pt}
\tablecaption{Detected Unidentified Lines}
\tablehead{
\colhead{Rest}&\multicolumn{2}{c}{BIMA Array}&\colhead{}&\multicolumn{2}{c}{12 meter telescope}&\colhead{Potential Identification} \\
\cline{2-3} \cline{5-6} \colhead{Frequency}&\colhead{I$_0$}&\colhead{$\Delta$v}&\colhead{}&\colhead{I$_0$}&\colhead{$\Delta$v}&\colhead{Transition, Frequency(MHz)}\\
\colhead{(MHz)}&\colhead{(Jy bm$^{-1}$)}&\colhead{(km s$^{-1}$)}&\colhead{}&\colhead{(Jy bm$^{-1}$)}&\colhead{(km s$^{-1}$)}&\colhead{and Species)}}
\startdata
85,907.7\tablenotemark{a}&0.7\tablenotemark{b}&3.4&& $<0.6$&\nodata&\\
85,910.5&1.0(4)&10.9(68)&& $<0.6$&\nodata&$9_{4,6}-8_{4,5}$, 85,910.564, C$_2$H$_3$CN $\nu$2=11(FJL)\\*
&&&&&&$9_{4,5}-8_{4,4}$, 85,910.632, C$_2$H$_3$CN $\nu$2=11(FJL)\\*
85,912.7\tablenotemark{a}&0.9\tablenotemark{b}&1.7&& $<0.6$&\nodata&\\
85,917.5&0.9(4)&6.9(36)&& $<0.6$&\nodata&$9_{6,*}-8_{6,*}$, 85,917.187, C$_2$H$_3$CN $\nu$2=11(FJL)\\
85,932.7\tablenotemark{a}&$<0.3$&\nodata&& 1.2(6)&3.9(28)&\\
85,935.1&1.0(4)&6.9(28)&& 2.1(4)&6.9(20)&\\
85,939.1&0.7(2)&10.5(54)&& $<0.6$&\nodata&\\
85,942.0\tablenotemark{a}&$<0.3$&\nodata&& -4.2(8)&3.1(6)&\\
85,945.9&0.9(4)&5.0(26)&& $<0.6$&\nodata&$9_{8,*}-8_{8,*}$, 85,945.968, C$_2$H$_3$CN $\nu$2=11(FJL)\\
85,970.6\tablenotemark{a}&0.7(6)&3.3(30)&& $<0.6$&\nodata&\\
85,973.2&0.4\tablenotemark{b}&5.7&& $<0.6$&\nodata&$13_{2,12}-12_{3,9}$AA, 85,973.249, (CH$_3$)$_2$O\citep{gron98}\\
85,987.4&1.9(4)&10.5(24)&& $<0.6$&\nodata&$9_{2,7}-8_{2,6}$, 85,987.176, C$_2$H$_3$CN $\nu$1=11(FJL)\\
86,010.0&0.8(4)&5.1(32)&& $<0.6$&\nodata&\\
86,024.3\tablenotemark{a}&$<0.3$&\nodata&& 4.1(8)&2.7(6)&\\
86,034.3&0.8(4)&5.1(32)&& $<0.6$&\nodata&\\
86,133.2&0.5(2)&10.0(72)&& $<0.6$&\nodata&$18_{4,14}^+-18_{3,16}^-$\tablenotemark{c}, 86,132.920, EtOH(JCP)\\*
&&&&&&$6_{3,3}-6{3,4}$ $J=11/2-11/2$, 86,129.608-86,133.000\tablenotemark{d}, c-CC$^{13}$CH\citep{jpl}\\*
86,148.0&0.5(4)&5.3(48)&& $<0.6$&\nodata&$6_{3,3}-6{3,4}$ $J=11/2-11/2$, 86,147.719-86150.854\tablenotemark{d}, c-CC$^{13}$CH\citep{jpl}\\
86,151.6&0.6(2)&17.3(140)&& $<0.6$&\nodata&\\
86,204.6&0.9(2)&5.4(20)&& $<0.6$&\nodata&\\
86,207.8&0.9(2)&7.4(28)&& $<0.6$&\nodata&\\
86,220.9&0.9(2)&4.7(24)&& 2.1(4)&14.9(40)&\\
86,226.5&0.7(2)&13.2(74)&& $<0.6$&\nodata&$2_{2,0}-2_{1,1}$EE, 86,226.727, (CH$_3$)$_2$O\citep{gron98}\\
86,239.6&$<0.2$&\nodata&& 1.7(4)&8.1(34)&\\
86,243.5&$<0.2$&\nodata&& 1.6(4)&10.6(46)&$2-1, \nu=1$, 86,243.440, SiO\tablenotemark{e} \citep{lovas03,snyder74}\\
86,248.2&0.8(2)&6.6(34)&& $<0.6$&\nodata&\\
86,255.2&0.6(2)&12.2(54)&& $<0.6$&\nodata&$9_{2,7}-8_{2,6}$, 86,254.848, C$_2$H$_3$CN $\nu$2=11(FJL)\\
86,263.5&$<0.2$&\nodata&& 1.4(4)&10.0(56)&\\
86,297.2&0.4(2)&5.0(38)&& $<0.6$&\nodata&\\
86,300.7&0.7(2)&12.5(40)&& $<0.6$&\nodata&$29-28$, 86,300.430, $^{30}$SiC$_4$(FJL)\\
86,386.3&0.9(4)&4.6(24)&& $<0.6$&\nodata&\\
86,389.2&-1.1(4)&2.2(12)&& -2.7(8)&2.4(8)&\\
86,395.9\tablenotemark{f}&0.9(2)&5.2(18)&& 1.9(6)&3.3(14)&\\
86,398.8&0.8(2)&7.4(26)&& $<0.6$&\nodata&\\
86,421.8&0.6(2)&6.7(32)&& $<0.6$&\nodata&\\
86,436.5\tablenotemark{a}&0.6(4)&1.9(12)&& $<0.6$&\nodata&\\
86,440.2&0.5(2)&17.8(76)&& $<0.6$&\nodata&\\
86,445.8&0.6(2)&9.9(46)&& $<0.6$&\nodata&\\
86,459.3&0.5(4)&5.2(46)&& $<0.6$&\nodata&$6_{3,3}-6{3,4}$ $J=13/2-13/2$, 86,458.768-86,459.831\tablenotemark{d}, c-CC$^{13}$CH\citep{jpl}\\
86,473.4\tablenotemark{f}&0.7(2)&8.9(26)&& $<0.6$&\nodata&\\
86,486.6&0.6\tablenotemark{b}&3.2&& 1.3(2)&24.2(72)&\\
86,498.2\tablenotemark{a}&0.7\tablenotemark{b}&1.9&& $<0.6$&\nodata&\\
86,536.6&0.6(2)&4.8(28)&& $<0.6$&\nodata&\\
86,543.7&$<0.2$&\nodata&& 2.4(6)&8.3(30)&\\
86,784.5&-1.4(6)&3.7(20)&& -2.2(8)&3.8(14)&\\
86,794.5&$<0.3$&\nodata&& -1.4(4)&5.1(20)&\\
86,803.6&-1.2(6)&2.2(18)&& -2.0(8)&2.2(12)&\\
86,812.3&$<0.3$&\nodata&& 1.6(2)&10.7(34)&\\
86,831.6&0.8\tablenotemark{b}&2.5&& 1.7\tablenotemark{b}&6.0&$13_3-14_2$A+, 86,831.480, CH$_3$SH(FJL)\\
86,915.5&0.8(2)&9.5(30)&& $<0.6$&\nodata&\\
86,944.5\tablenotemark{a}&1.0(4)&3.7(20)&& $<0.6$&\nodata&\\
86,956.1\tablenotemark{a}&0.7(2)&3.6(12)&& $<0.6$&\nodata&\\
86,957.8\tablenotemark{a}&0.8(2)&2.2(10)&& $<0.6$&\nodata&\\
86,976.0&1.3(4)&4.4(16)&& 1.5(6)&3.9(20)&\\
86,979.3\tablenotemark{f}&1.9(2)&9.5(18)&& 1.6(4)&10.1(32)&$10_{1,10}-9_{1,9}$A, 86,978.865, EtCN(JCP)\\*
&&&&&&$10_{1,9}-9_{1,8}$E, 86,978.884, EtCN(JCP)\\*
87,008.3&0.5(2)&14.5(40)&& $<0.6$&\nodata&\\
87,016.5&0.5(2)&8.5(28)&& $<0.6$&\nodata&$4_{3,2}-3_{3,1}$, 87,016.896, HCOOD\citep{jpl}\\
87,027.4&0.7(2)&6.7(24)&& $<0.6$&\nodata&$5_{2,3}^--4_{2,2}^-$\tablenotemark{c}, 87,027.093, EtOH(JCP)\\
89,731.9&0.9(4)&7.1(36)&& $<0.7$&\nodata&\\
89,739.6&4.1(2)&22.6(20)&& 2.9(4)&15.7(22)&$10_{4,6}-9_{4,5}$A, 89,739.381, EtCN(JCP)\\*
&&&&&&$10_{4,6}-9_{4,5}$E, 89,739.935, EtCN(JCP)\\*
89,754.9&1.6(8)&4.8(26)&& 1.2(6)&5.3(32)&\\
89,767.3&1.7(2)&32.6(54)&& $<0.7$&\nodata&\\
89,791.5&0.6(2)&18.4(82)&& $<0.7$&\nodata&\\
89,814.7&0.8(2)&11.9(40)&& 1.1(4)&8.8(32)&$33_{1,32}-33_{1,33}$, 89,814.7342-89,816.2485\tablenotemark{g}, HNCO\citep{jpl}\\
89,829.6&0.8(4)&5.9(26)&& $<0.7$&\nodata&\\
89,897.3&0.3(2)&15.4(14)&& $<0.7$&\nodata&\\
89,902.3&0.5(4)&5.4(42)&& $<0.7$&\nodata&\\
89,923.8&0.4(2)&9.1(60)&& $<0.7$&\nodata&\\
89,959.4\tablenotemark{a}&0.8(4)&3.4(20)&& $<0.7$&\nodata&\\
89,989.7&0.8(2)&10.3(36)&& $<0.7$&\nodata&\\
89,993.1&0.8(2)&7.5(26)&& $<0.7$&\nodata&$33_{1,32}-33_{1,33}$, 89,992.9883-89,993.4035\tablenotemark{g}, HN$^{13}$CO\citep{jpl}\\
90,004.8&0.6(2)&5.5(30)&& $<0.7$&\nodata&\\
90,007.7&0.7(2)&8.8(38)&& $<0.7$&\nodata&$13_{0,13}^+-13_{1,13}^+$\tablenotemark{c}, 90,007.693, EtOH(JCP)\\
90,033.9&1.0(2)&11.6(50)&& $<0.7$&\nodata&\\
90,037.8\tablenotemark{f}&1.7(2)&12.1(40)&& 2.2(4)&8.2(20)&\\
90,069.1&0.5(2)&6.3(34)&& $<0.7$&\nodata&\\
90,078.3&0.8(2)&18.5(38)&& $<0.7$&\nodata&$17_{2,15}-17_{2,16}$, 90,077.914, HCOOH\citep{jpl}\\
90,128.5&0.5(2)&4.8(32)&& $<0.7$&\nodata&\\
90,142.9&0.8(4)&4.1(26)&& $<0.7$&\nodata&\\
90,153.8&$<0.2$&\nodata&& 2.3(6)&5.3(22)&\\
90,166.9&1.2(6)&6.8(32)&& $<0.7$&\nodata&\\
90,174.5&0.6(2)&11.8(62)&& $<0.7$&\nodata&\\
90,179.1&0.7(2)&7.4(36)&& $<0.7$&\nodata&$24_{4,21}^+-23_{5,19}^-$\tablenotemark{c}, 91,178.955, EtOH(JCP)\\
90,203.4\tablenotemark{f}&0.7(2)&13.2(36)&& $<0.7$&\nodata&\\
90,224.5&1.3(4)&1.6(6)&& 2.2(8)&2.2(10)&\\
90,304.8&0.6(4)&4.1(34)&& $<0.7$&\nodata&\\
90,307.9&0.6(2)&12.0(66)&& $<0.7$&\nodata&$23_{3,21}-22_{4,19}$, 90,307.778, C$_2$H$_5$OOCH\citep{jpl}\\
106,295.4&1.0(4)&9.1(56)&& $<0.6$&\nodata&\\
106,322.0&1.0(2)&22.9(84)&& $<0.6$&\nodata&$18_{2,16}^+-18_{1,17}^+$\tablenotemark{c}, 106,322.066, EtOH(JCP)\\
106,353.0&0.9(2)&17.3(66)&& $<0.6$&\nodata&$22_{3,19}-21_{4,18}$, 106,352.655-106,353.379\tablenotemark{d}, SO$^{17}$O(FJL)\\*
&&&&&& $32_{3,29}-33_{2,32}$, 106,351.742-106,353.062\tablenotemark{d}, SO$^{17}$O(FJL)\\*
&&&&&&$11_{1,10}-10_{2,9}$, 106,352.528-106,352.878\tablenotemark{d}, DNO\citep{jpl}\\*
106,371.5\tablenotemark{a}&1.9(6)&3.9(16)&& $<0.6$&\nodata&\\
106,411.6&1.4(4)&7.3(30)&& $<0.6$&\nodata&$5_{2,3}-4_{2,2}$, 106,411.010-106,411.461\tablenotemark{d}, NH$_2^{13}$CHO(FJL)\\*
&&&&&& $35_{22,14}-35_{21,15}$AA, 106,411.273, (CH$_3$)$_2$CO \citep{gron02,snyder02}\\*
106,448.9&3.4(4)&7.7(14)&& 1.8(6)&5.5(20)&\\
106,468.5&0.8(6)&10.6(74)&& $<0.6$&\nodata&\\
106,484.8\tablenotemark{a}&1.5(8)&1.9(10)&& $<0.6$&\nodata&$5_{1,5}-4_{1,4}$, 106,484.610, HO$^{13}$CO$^+$\citep{jpl}\\*
&&&&&&$15_{6,10}^--16_{5,12}^+$\tablenotemark{c}, 106,484.991, EtOH(JCP)\\
106,487.2&1.1(2)&13.9(46)&& $<0.6$&\nodata&\\
106,510.7&1.5(4)&5.7(22)&& $<0.6$&\nodata&\\
106,514.5&1.8(4)&5.9(18)&& $<0.6$&\nodata&\\
106,559.8&0.9(4)&5.8(38)&& $<0.6$&\nodata&\\
106,629.9&1.2(4)&6.6(30)&& $<0.6$&\nodata&\\
106,638.4&4.3(18)&6.6(24)&& 4.0(8)&6.6(14)&$4_5-3_4$, 106,638.559, HC$_2^{15}$N(FJL)\\*
&&&&&&$11_{2,12}-10_{2,11}$, 106,638.577, C$_4$H\citep{cologne}\\*
106,654.8&2.8(4)&6.4(14)&& 1.7(4)&12.2(42)&\\
106,706.7&1.2(6)&5.4(32)&& $<0.6$&\nodata&\\
106,714.3&1.0(6)&4.5(36)&& $<0.6$&\nodata&\\
106,753.1&1.5(4)&8.8(26)&& 2.2(6)&6.1(18)&$11_{1,10}-10_{1,9}$ vt=2, 106,753.070, C$_2$H$_3$CN\citep{jpl}\\
106,763.3\tablenotemark{a}&1.3(8)&2.2(16)&& $<0.6$&\nodata&\\
106,767.4&1.0(4)&5.2(28)&& $<0.6$&\nodata&$6_{1,5}^--5_{1,4}^-$\tablenotemark{c}, 106,767.176, EtOH(JCP)\\
108,207.3&1.8(16)&4.3(46)&& $<0.9$&\nodata&$2_{2,0}-3_{1,3}$, 108,207.399, VyCN\citep{jpl}\\
108,223.7\tablenotemark{a}&2.1(14)&3.9(32)&& $<0.9$&\nodata&\\
108,379.8&1.9(12)&3.5(24)&& 2.3(6)&6.6(20)&\\
108,434.9&1.6(4)&5.0(12)&& 1.6(8)&7.7(40)&$14_{2,12}-14_{1,13}$EE, 108,434.511, (CH$_3$)$_2$CO\citep{gron02,snyder02}\\*
&&&&&&$26-25, J=49/2-49/2$, 108,435.7053, l-HC$_4$N\citep{cologne}\\*
108,507.7&1.3(4)&9.1(68)&& $<0.9$&\nodata&\\
108,511.2&1.3(6)&7.4(62)&& $<0.9$&\nodata&$31_{3,29}-30_{4,26}$, 108,511.262-108,511.299\tablenotemark{d}, $^{33}$SO$_2$\citep{cologne}\\
108,620.7\tablenotemark{a}&2.2(8)&1.8(10)&& $<0.9$&\nodata&\\
109,943.3&2.0(4)&5.0(16)&& $<0.6$&\nodata&\\
109,959.6&3.5(4)&10.5(16)&& 2.1(6)&8.1(44)&$33_{1,33}-34_{0,34}$, 109,959.100, HNCO\citep{winn76}\\*
&&&&&& $22_{1,21}-22_{1,22}$, 109,958.699, EtCN\citep{lovas82}\\*
&&&&&& $16_{8,9}-15_{8,8}$, 109,960.360, NH$_2$CH$_2$COOH-I (glycine)\tablenotemark{h} (FJL)\\*
109,972.0&0.9(6)&4.1(30)&& $<0.6$&\nodata&\\
110,028.6\tablenotemark{a}&$<0.5$&\nodata&& 2.4(16)&1.3(10)&\\
110,036.2&3.5\tablenotemark{b}&3.3&& 1.5(8)&5.9(36)&\\
110,079.7&0.9(4)&6.7(40)&& $<0.6$&\nodata&\\
110,085.6&2.0(4)&13.3(38)&& 1.9(8)&6.3(26)&$4_{2,3}-3_{1,2}$A vt=1, 110,085.727, EtCN(JCP)\\
110,090.2&1.2(6)&4.7(28)&& $<0.6$&\nodata&$6_{5,2}-5_{4,1}$AE, 110,089.569, (CH$_3$)$_2$CO\citep{gron02,snyder02}\\*
&&&&&&$7_{5,3}-6_{4,2}$, 110,089.998, NH$_2$CH$_2$COOH-I\tablenotemark{i} (FJL)\\*
&&&&&&$4_{2,2}-3_{1,3}$E vt=1, 110,090.695, EtCN(JCP)\\*
110,098.0&1.2(4)&10.8(28)&& $<0.6$&\nodata&\\
110,104.7&1.7(4)&10.8(38)&& $<0.6$&\nodata&$4_{4,0}-3_{2,1}$, 110,104.847, NH$_2$CH$_2$COOH-I\tablenotemark{i} (FJL)\\
110,143.3&2.5\tablenotemark{b}&16.7&& $<0.6$&\nodata&\\
110,150.0&2.2(6)&7.8(26)&& $<0.6$&\nodata&$9_{2,7}-8_{3,6}$EA, 110,149.102, (CH$_3$)$_2$CO\citep{gron02,snyder02}\\*
&&&&&& $9_{2,7}-8_{3,6}$AE,  110,148.838, (CH$_3$)$_2$CO\citep{gron02,snyder02}\\*
110,158.2&$<1.0$&\nodata&& 5.1\tablenotemark{j}&31.8&$26_{21,6}-26_{20,7}$EA, 110,158.590, (CH$_3$)$_2$CO\citep{gron02,snyder02}\\
&&&&3.6\tablenotemark{j}&1.9&\\
110,161.0\tablenotemark{a}&$<1.0$&\nodata&& 3.5(14)&2.6(12)&\\
110,164.6\tablenotemark{a}&$<1.0$&\nodata&& 4.5(12)&3.1(12)&$5_{1,4}-4_{1,3}$, 110,164.245, HNCO $\nu_6$\tablenotemark{k} \citep{lovas03,wyr99}\\
110,213.0&\tablenotemark{l}&\nodata&& \tablenotemark{l}&\nodata&\\
110,216.4&$<0.8$&\nodata&& \tablenotemark{l}&\nodata&$5_{4,1}-4_{4,0}$, 110,216.85, DCOOH\citep{will80}\\*
&&&&&& $5_{4,2}-4_{4,1}$, 110,216.64, DCOOH\citep{will80}\\*
110,222.2&\tablenotemark{l}&\nodata&& \tablenotemark{l}&\nodata&\\
110,233.6&\tablenotemark{l}&\nodata&& \tablenotemark{l}&\nodata&\\
110,239.9\tablenotemark{f}&\tablenotemark{l}&\nodata&& \tablenotemark{l}&\nodata&\\
110,246.6&$<0.8$&\nodata&& \tablenotemark{l}&\nodata&\\
110,253.0&\tablenotemark{l}&\nodata&& $<0.6$&\nodata&$15_{15}-14_{14}$, 110,252.932, C$^{13}$CS\citep{jpl}\\
110,255.5&$<0.5$&\nodata&& \tablenotemark{l}&\nodata&$4_{4,1}-5_{3,2}$AA, 110,256.338, (CH$_3$)$_2$O\citep{gron98}\\
110,263.8&\tablenotemark{l}&\nodata&& \tablenotemark{l}&\nodata&\\
110,418.4&2.9(10)&9.4(46)&& $<0.6$&\nodata&$22_{20,2}-22_{19,3}$EA, 110,417.969, (CH$_3$)$_2$CO\citep{gron02,snyder02}\\
110,421.4\tablenotemark{a}&2.8(18)&2.9(24)&& $<0.6$&\nodata&$36-35$, 110,420.724, SiC$_4$\citep{jpl}\\
110,446.8&2.6(8)&6.4(26)&& $<0.6$&\nodata&$9_{8,1}-8_{8,0}$E, 110,447.032, MeF\citep{oest99}\\*
110,447.9&$<0.97$&\nodata&& 2.6(10)&4.9(20)&$14_{3,12}-14_{2,13}$A, 110,447.595, EtCN(JCP)\\

\enddata
\tablenotetext{a}{This feature is a questionable detection because its line width is $<$4 km s$^{-1}$; hence it may be due to interference.}
\tablenotetext{b}{Intensity and line width were approximated because the least squares Gaussian fitting did not give a satisfactory fit.}
\tablenotetext{c}{For EtOH a ``+''= gauche+ and a ``-''= gauche- states.}
\tablenotetext{d}{Large frequency spread due to hyperfine components.}
\tablenotetext{e}{The transition has been previously reported as a masing transition in Orion by \citet{snyder74}. However due to its large line width combined with its very high E$_u$ ($\sim$1760 K) it is unlikely that this identification is correct for this source.}
\tablenotetext{f}{Previously detected U line.}
\tablenotetext{g}{Large frequency spread due to multiple $J$ components.}
\tablenotetext{h}{It is not likely that this line is from glycine since it is detected strongly with both instruments and glycine is expected to be very compact in distribution.}
\tablenotetext{i}{It is not likely that this line is from glycine since this transition has an unfavorable line strength.}
\tablenotetext{j}{Multiple components were necessary to get an approximate least squares Gaussian fit. Each component is listed on a separate line.}
\tablenotetext{k}{This transition has been previously reported as a HNCO transition toward G10.47+0.03 by \citet{wyr99}. However due to its narrow line width we are reporting it as a questionable detection for Sgr~B2(N-LMH).}
\tablenotetext{l}{Multiple emission and absorption components detected.}
\label{tab:U}
\end{deluxetable}

\begin{deluxetable}{lcccccccc}
\tabletypesize{\scriptsize}
\tablecolumns{9}
\tablewidth{0pt}
\tablecaption{Comparison of Line Statistics}
\tablehead{
\colhead{} & \multicolumn{2}{c}{Identified} & \colhead{} & \multicolumn{2}{c}{Unidentified}& \colhead{} & \multicolumn{2}{c}{Total} \\
\cline{2-3} \cline{5-6} \cline{8-9}\\
\colhead{Source} & \colhead{Lines} & \colhead{Density\tablenotemark{a}} & \colhead{} & \colhead{Lines} & \colhead{Density\tablenotemark{a}} & \colhead{} & \colhead{Lines} & \colhead{Density\tablenotemark{a}}}
\startdata
\citet{cummins86} & 49 & 1.36 & & 3 & 0.08 & & 52 & 1.44 \\
\citet{turner89} & 57 & 1.58 & & 10 & 0.28 & & 67 & 1.86 \\
\citet{lovas03} & 151 & 4.19 & & 38 & 1.06 & & 189 & 5.25 \\
This survey & 98 & 2.72 & & 120 & 3.34 & & 218 & 6.06\\
\enddata
\tablenotetext{a}{Lines per 100 MHz.}
\label{tab:stats}
\end{deluxetable}

\end{document}